\documentclass[a4paper,fleqn,usenatbib]{mnras}

\usepackage[T1]{fontenc}
\usepackage{ae,aecompl}

\usepackage{graphicx}	
\usepackage{caption}
\usepackage{amsmath}	
\usepackage{amssymb}	

\title[Nonlinear properties of IGR J17091-3624]{Correlating nonlinear time series and spectral properties of IGR J17091-3624: Is it similar to GRS~1915+105?}

\author[Adegoke, Mukhopadhyay \& Misra]{
Oluwashina Adegoke,$^{1}$\thanks{E-mail: oluwashinaa@iisc.ac.in}
Banibrata Mukhopadhyay,$^{1}$\thanks{E-mail: bm@iisc.ac.in}
Ranjeev Misra$^{2}$\thanks{E-mail: rmisra@iucaa.in}
\\
$^{1}$Department of Physics, Indian Institute of Science, Bangalore 560012, India\\
$^{2}$Inter-University Centre for Astronomy and Astrophysics,
Post Bag 4, Ganeshkhind, Pune 411007, India\\
}

\date{Accepted 2019 December 30. Received 2019 December 9; in original form 2019 August 21}

\pubyear{2019}

\begin{document}
\label{firstpage}
\pagerange{\pageref{firstpage}--\pageref{lastpage}}
\maketitle

\begin{abstract}
We explore the nonlinear properties of \textrm{IGR~J17091-3624} in the line of the underlying behaviour for
\textrm{GRS~1915+105}, following correlation integral method. We find that while the latter is known to reveal the combination of fractal (or even chaotic) and
stochastic behaviours depending on its temporal class, the former mostly shows stochastic behaviour.
Therefore, although several observations argue \textrm{IGR~J17091-3624} to be similar to 
\textrm{GRS~1915+105} with common temporal classes between them, underlying nonlinear time series
analyses offer differently. Nevertheless, the Poisson noise to $rms$ variation value for 
 \textrm{IGR~J17091-3624} turns out to be high, arguing them to be 
Poisson noise dominated and hence may plausibly lead to suppression of the nonlinear 
character, if any. Indeed, it is a very faint source compared to \textrm{GRS~1915+105}. 
However, by increasing time bin some of the temporal classes of \textrm{IGR~J17091-3624} 
show deviation from 
stochasticity, indicating plausibility of higher fractal dimension.  Along with spectral analysis, overall \textrm{IGR~J17091-3624} argues 
to reveal three different accretion classes: slim, Keplerian and advective accretion discs.

\end{abstract}




\section{Introduction}
On timescales of seconds to minutes, the three sources, \textrm{GRS~1915+105}, \textrm{IGR~J17091-3624} and \textrm{MXB~1730-335}, have been shown to exhibit a variety of peculiar variability properties in their lightcurves not normally seen in other low mass X-ray binaries (LMXBs). In the black hole X-ray binaries (BHXBs) \textrm{GRS~1915+105} and \textrm{IGR~J17091-3624}, these properties are categorized into variability classes. While most LMXBs transit between quiescence and outbursts (including \textrm{IGR~J17091-3624}), \textrm{GRS~1915+105} has been in outburst since it was discovered in 1992 \citep{1992IAUC.5590....2C}. 

\textrm{IGR~J17091-3624}, on the other hand, is an LMXB discovered during an outburst by \textit{INTEGRAL} in 2003 \citep{2003ATel..149....1K}. It again went into outburst in 2011 \citep{2011ATel.3148....1K} and was observed by \textit{RXTE}, \textit{Swift}, \textit{XMM-Newton}, \textit{Chandra} and \textit{Suzaku} and most recently in 2016 (\citealt{2016ATel.8742....1M, 2016ATel.8761....1G}). During the long duration of the $2011$ outburst, analyses of the lightcurves show \textrm{GRS~1915+105} like properties in several of the individual observations \citep{2011ApJ...742L..17A, 2012ApJ...747L...4A, 2014ApJ...783..141P, 2017MNRAS.468.4748C}. While 
\textrm{GRS~1915+105} reveals 12 distinct temporal classes \citep{2000A&A...355..271B}, \textrm{IGR~J17091-3624}
is also found to exhibit 9 temporal classes \citep{2011ApJ...742L..17A,2017MNRAS.468.4748C}.
Nevertheless, at $2-25\,\mathrm{keV}$, \textrm{IGR~J17091-3624} is about a factor of 20 fainter than \textrm{GRS~1915+105} implying plausibly that \textrm{IGR~J17091-3624} is at a larger distance, and/or hosts a smaller black hole or has a lower accretion rate $\dot{m}$ than \textrm{GRS~1915+105}. 

The time series of \textrm{GRS~1915+105} was repeatedly found to exhibit, sometimes fractal (or even
chaotic) and sometimes stochastic behaviours \citep{2004ApJ...609..313M, 2004AIPC..714...48M, 2006ApJ...643.1114M} based on the correlation integral (CI)
calculation, depending on its
temporal class \citep{2000A&A...355..271B}. In an attempt to understanding the bursting behaviour seen in \textrm{GRS~1915+105}, 
\citet{2014Ap&SS.352..699M}
 and
 \citet{2017IJNLM..88..142A}
 considered a simple mathematical framework composed of nonlinear differential equations. They concluded from their modelling that the complex behaviour seen in the source may be regulated mainly by a single nonlinear oscillator that is driven by a unique parameter -- plausibly the outer mass accretion rate -- whose changes may be responsible for the variability classes. Recently, based on correlating temporal (time series) 
and spectral properties, the source has been argued to exhibit different accretion classes:
advection dominated accretion \citep{1994ApJ...428L..13N}, general advective accretion \citep{2010MNRAS.402..961R}, Keplerian accretion disc \citep{1973A&A....24..337S}, and
slim disc \citep{1988ApJ...332..646A} flows, switching from one to another with time \citep{2018MNRAS.476.1581A}. As 
\textrm{IGR~J17091-3624} has already been found to have similarities in several observations
with \textrm{GRS~1915+105}, naturally the question arises if it also exhibits the combination of 
fractal/chaotic and stochastic natures in classes and also similar temporal and spectral correlations
arguing for 4 accretion classes, as found in \textrm{GRS~1915+105}? This we plan to explore here. 

The plan of the paper is the following. In the next section, we give an overview of the CI method. We discuss the observed
data under consideration from three satellites in \S 3. Subsequently, we explain various
results by analyzing data in \S 4. Significance of observed stochasticity in order to
conclude from such analyses is discussed in \S 5. Finally we end with a 
conclusion along with discussions in \S 6.

\section{Correlation Integral method}


The delay embedding technique of \citet{1983PhRvA..28.2591G} is one of the standard methods used to reconstruct the dynamics of a nonlinear system from a time series. The detailed technique in the astrophysical context was already
discussed earlier (\citealt{2010ApJ...708..862K}), here we briefly recall them
for completeness.
 It requires constructing a CI defined as the probability that two points in phase space are closer together than a distance $r$. It is normally required that the dynamics be constructed for different embedding dimensions $M$, since the number of equations or variables describing the system is not known a priori.
  In using this method, vectors of length $M$ are created from the time series $y(t_{i})$ using a delay time $\tau$ such that
\begin{eqnarray} 
X(t_{i})=(y(t_{i}), y(t_{i}+\tau), y(t_{i}+2\tau),...,y(t_{i}+(M-1)\tau))
\end{eqnarray}
for the $i$-th vector. 

As a condition, the times $t_{i}$ should be evenly spaced and should be chosen such that the number of data 
points is sufficiently large. 
More so, the choice of $\tau$ should be such that any two components of vectors are not correlated.

The quantity $D_{2}$ known as correlation dimension provides a quantitative picture of this technique. Computationally, it involves choosing a 
large number of points in the reconstructed dynamics as centres and then computing CI $C_{M}(r)$, which is the number of points that are within a distance $r$ from the centre averaged over all the centres, written as
\begin{eqnarray}
C_{M}(r)=\frac{1}{N({N_{c}-1})}\mathop{\sum^{N}\sum^{N_{c}}}_{i=1\ j\neq i\ j=1}{H(r-|{x_{i}-x_{j}}|)},
\end{eqnarray}
where $N$ and $N_{c}$ are the number of points and the number of centres respectively, $x_{i}$ and $x_{j}$ are the 
reconstructed vectors and $H$ is the Heaviside step function. 

The correlation dimension $D_{2}$ being just a 
scaling index of the variation of $C_{M}(r)$ with $r$ is expressed as
\begin{eqnarray}
D_{2}=\mathop {\lim }\limits_{r \to 0}\left(\frac{d\rm{log}C_{M}(r)}{d\rm{logr}}\right).
\end{eqnarray}
The effective number of differential equations governing the dynamics of the system can be inferred from the value of $D_{2}$ in principle. The linear part of the plot of $\rm{log}[C_M(r)]$ against $\rm{log}[r]$ can be used to estimate $D_{2}$. The nonlinear dynamical properties of the system is revealed from the variation of $D_{2}$ with $M$. If for all $M$, $D_{2}\approx M$ then the system is stochastic. On the other hand, the system 
is deterministic if initially $D_{2}$ increases linearly with $M$ until it reaches a certain value and saturates. 
This saturated value of $D_{2}$ is then taken to be the fractal/correlation dimension of the system, plausibly a signature of chaos (subject to the confirmation by other methods). While the saturated $D_{2}$ (i.e. $D_{2}^{sat}$) gives a quantitative measure of the dynamics, the corresponding critical dimension $M_{cr}$ is a measure of the number of equations required to describe the behaviour of the system. 

\begin{table*}
\small
\caption{IGR~J17091-3624: Basic flow classes}
\label{tab:tab1}
\resizebox{\linewidth}{!}{
\begin{tabular}{cccccccccccccccccccccc}\\
\hline
\hline
ObsID & class & GRS~1915-like class &  behaviour & \textit{diskbb} & PL & GA &  SI  &
$\chi^{2}/dof$ & state & \textit{diskbb} $T_{in}$  & $F\,\times10^{-10}$ \\
\hline
\hline
96420-01-01-00 & I & $\chi$  & S & $10.0$ & $89.1$ & $0.9$ & $2.26^{+0.02}_{-0.02}$ & $1.12\,(48/43)$ & PD & $1.11\pm{0.05}$ & $11.949$\\
\hline
96420-01-11-00 & II & $\phi$ &  S & $18.5$ & $81.5$ & $-$ & $2.33^{+0.04}_{-0.04}$ & $1.04\,(47/45)$ & PD & $1.13\pm{0.02}$ & $6.238$\\
\hline
96420-01-04-01 & III & $\nu$ & S & $38.0$  & $62.0$ & $-$ & $2.25^{+0.07}_{-0.07}$ & $0.82\,(37/45)$ & PD & $1.20\pm{0.01}$ & $9.730$\\
\hline
96420-01-05-00 & IV & $\rho$ &  S & $41.6$ & $58.0$ & $0.4$ & $2.34^{+0.05}_{-0.05}$ & $1.06\,(45/43)$ & PD & $1.20\pm{0.01}$ & $9.676$\\
\hline
96420-01-06-03 & V & $\mu$ & S/NS & $54.4$ & $45.6$ & $-$ & $2.55^{+0.07}_{-0.07}$ & $1.28\,(58/45)$ & D-P & $1.43\pm{0.01}$ & $9.313$  \\
\hline
96420-01-09-00 & VI & $\lambda$ & S & $65.8$ & $34.2$ & $-$ & $2.55^{+0.09}_{-0.09}$ & $1.37\,(61/45)$ & DD & $1.85\pm{0.03}$ & $10.099$\\
\hline
96420-01-18-05 & VII & $none$ & S & $63.4$ & $36.6$ & $-$ & $2.61^{+0.15}_{-0.15}$ & $0.65\,(29/45)$ & DD & $1.50\pm{0.03}$ & $8.603$ \\
\hline
96420-01-19-03 & VIII & $none$ & S/NS & $73.4$ & $26.4$ & $0.2$ & $2.29^{+0.12}_{-0.12}$ & $1.30\,(56/43)$ & DD & $1.78\pm{0.02}$ & $10.596$\\
\hline
96420-01-35-02 & IX & $\gamma$ &  S & $68.2$ & $31.8$ & $-$ & $2.52^{+0.08}_{-0.08}$ & $1.58\,(71/45)$ & DD & $2.07\pm{0.03}$ & $11.409$\\
\hline
\hline
\hline
\end{tabular}
}
{Columns:- 
1: \textit{RXTE} observational identification number (ObsID).
2: Temporal class.
3: GRS~1915+105-like temporal class.
4: The behaviour of the system (F: low correlation/fractal dimension; S: Poisson noise like stochastic).
5: $\%$ of multi-colour blackbody component.
6: $\%$ of powerlaw component.
7: $\%$ Gaussian line component (XSPEC model gauss).
8: Powerlaw photon spectral index.
9: Reduced $\chi^2$.
10: Spectral state (DD: disc dominated; D-P: disc-powerlaw contributed; PD: 
powerlaw dominated).
11: \textit{diskbb} temperature in units of $keV$.
12: Total flux in the $3-25$ keV in units of $erg\,cm^{-2}\,s^{-1}$ 
}
\end{table*}

\begin{figure*}
\centering
 \begin{tabular}{@{}cc@{}}
  \includegraphics[scale=0.20, angle=-90]{plot2.eps} 
  \includegraphics[scale=0.20, angle=-90]{plot5.eps} 
  \includegraphics[scale=0.20, angle=-90]{plot7.eps} \\
  \\
  \includegraphics[scale=0.45, angle=0]{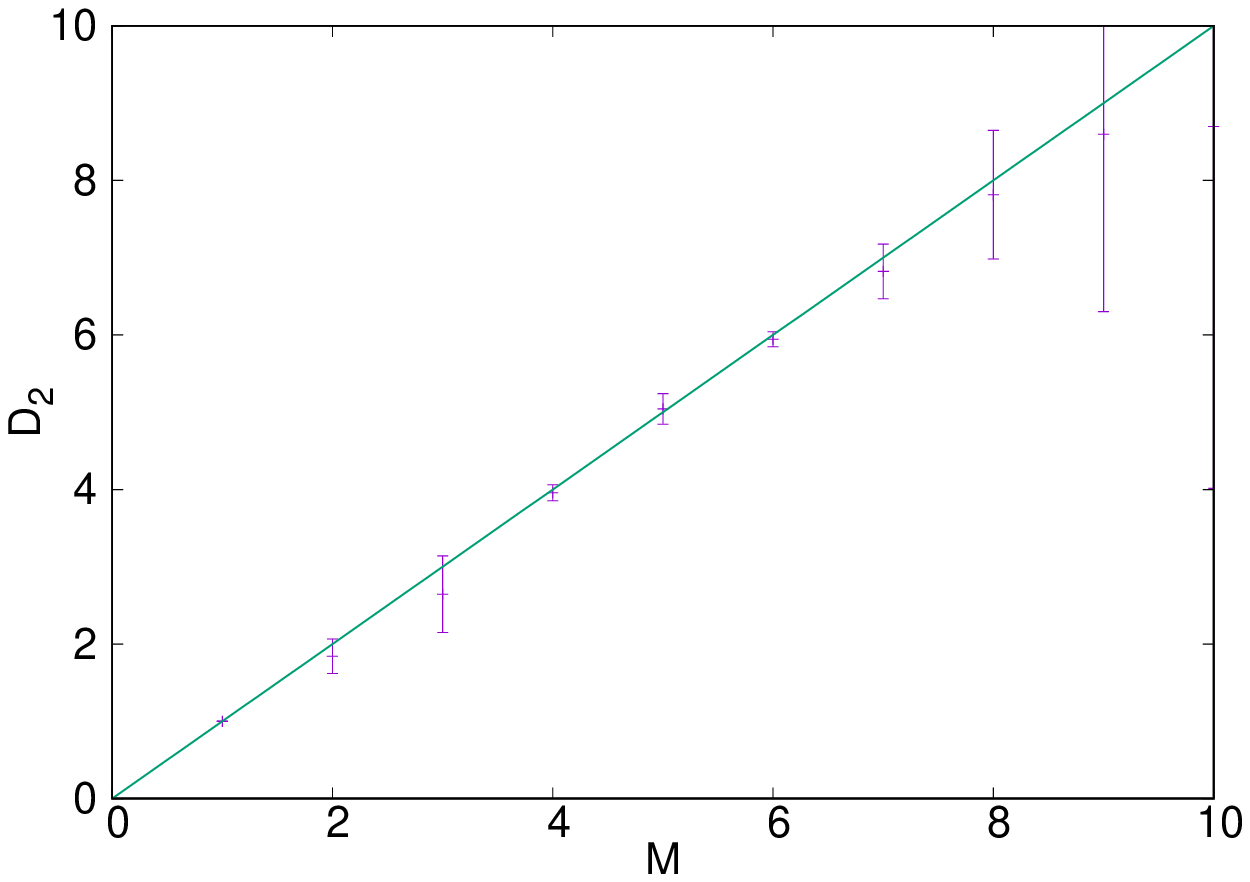} 
  \includegraphics[scale=0.45, angle=0]{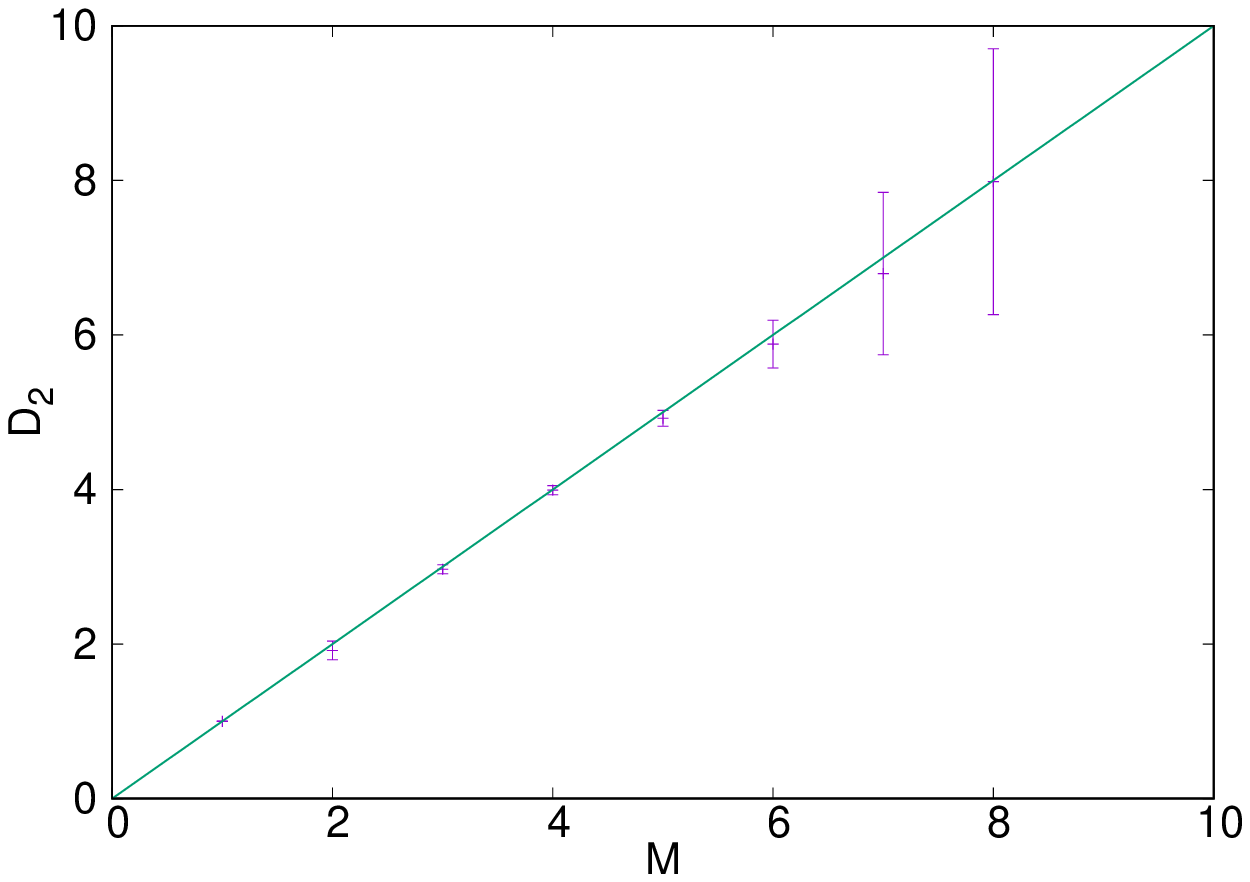} 
  \includegraphics[scale=0.45, angle=0]{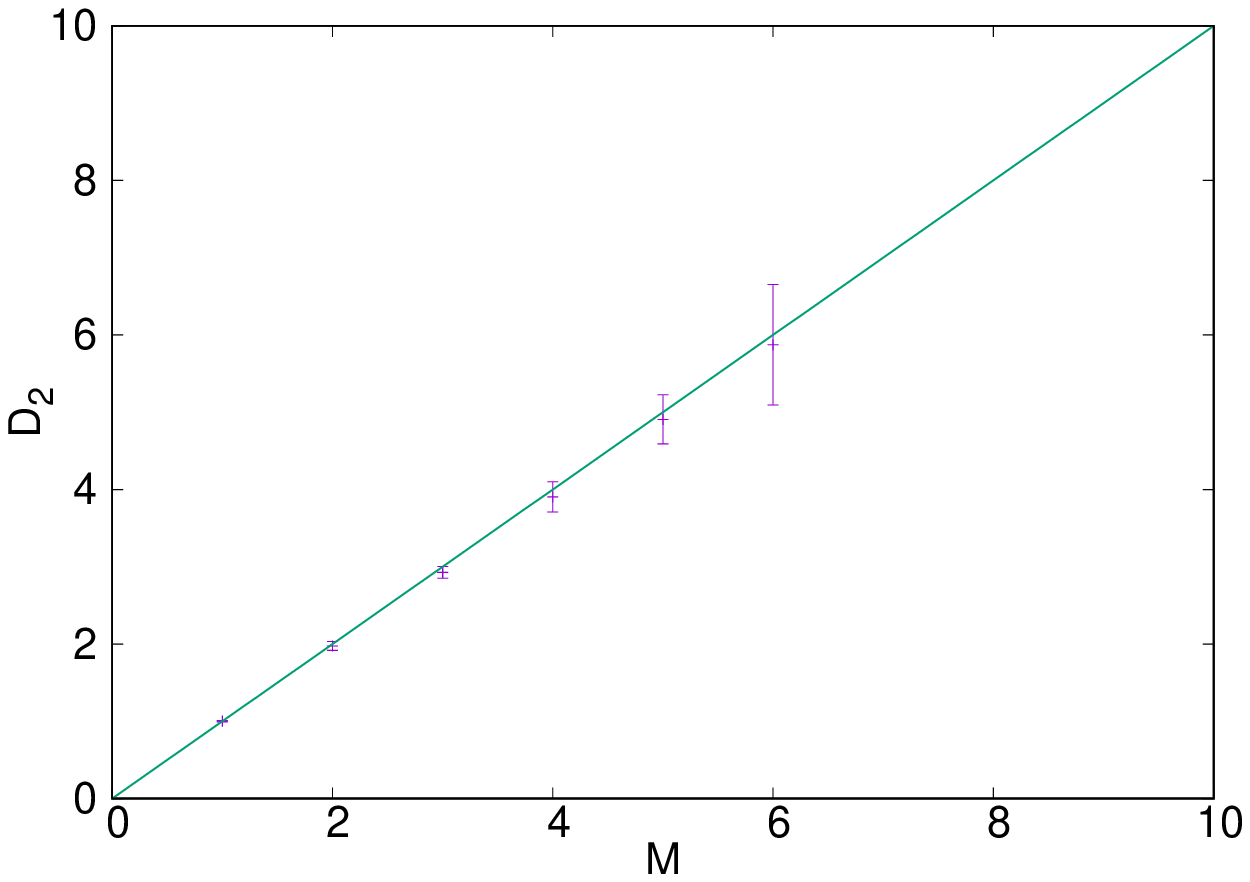} \\
 \end{tabular}
\caption{Top panels (\textit{left to right}): Unfolded spectra for \textit{RXTE} ObsIDs $96420-01-11-00$ (Class II), $96420-01-06-03$ (Class V) and $96420-01-18-05$ (Class VII) of IGR~J17091-3624. The curves with square and circle symbols represent respectively \textit{diskbb} and powerlaw contributions and the topmost solid curve in each panel represents the overall spectrum. 
Bottom panels (\textit{left to right}): Variation of correlation dimension as a function of embedding dimension with errors for \textit{RXTE} ObsIDs $96420-01-11-00$ (Class II), $96420-01-06-03$ (Class V) and $96420-01-18-05$ (Class VII) of IGR~J17091-3624. The solid diagonal line in each panel represents the ideal stochastic curve. See Tables \ref{tab:tab1} and \ref{tab:tab4} for other details.
}
\label{fig:fig1}
\end{figure*}
\begin{figure*}
\centering
 \begin{tabular}{@{}cc@{}}
  \includegraphics[scale=0.20, angle=-90]{xmm06.eps} &
  \includegraphics[scale=0.20, angle=-90]{xmm07.eps}
  \\
  \includegraphics[scale=0.45, angle=0]{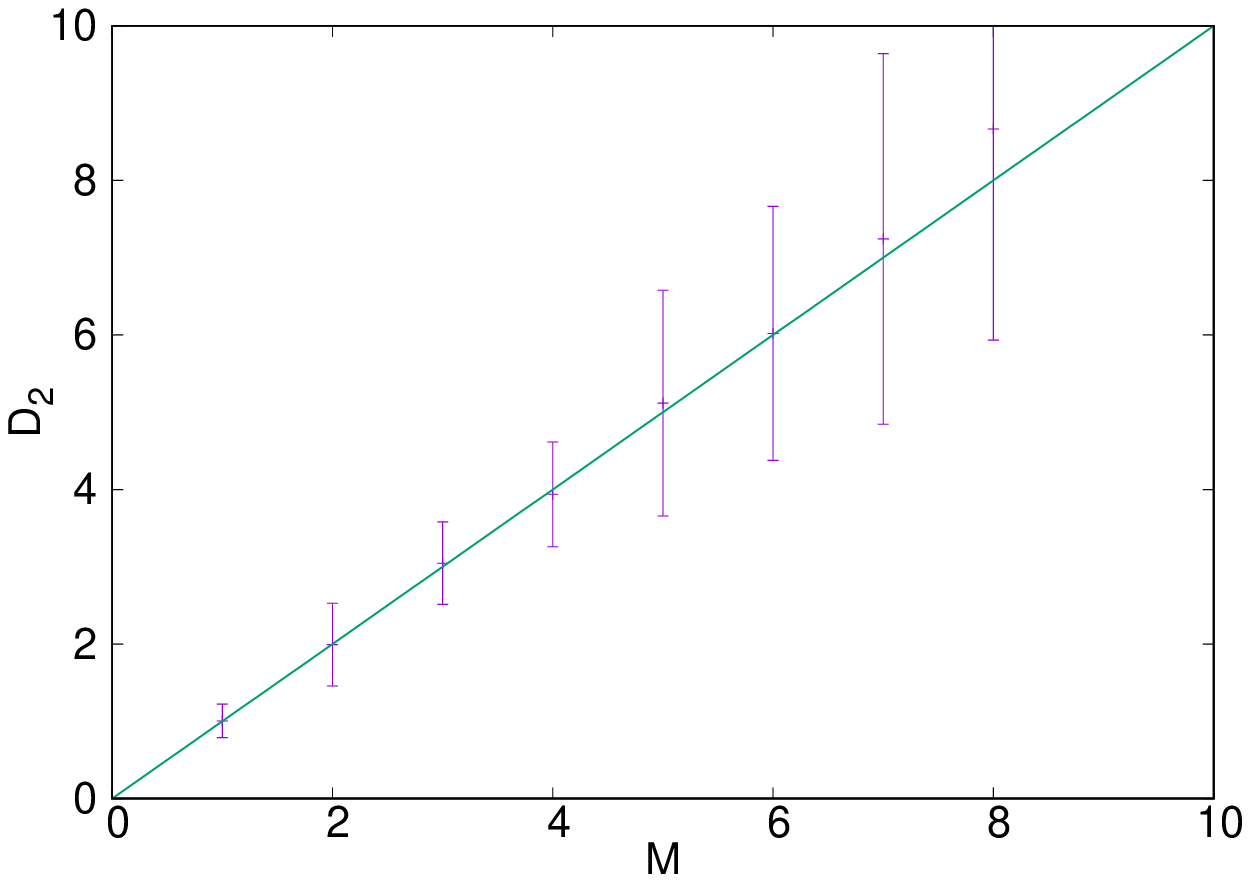} &
  \includegraphics[scale=0.45, angle=0]{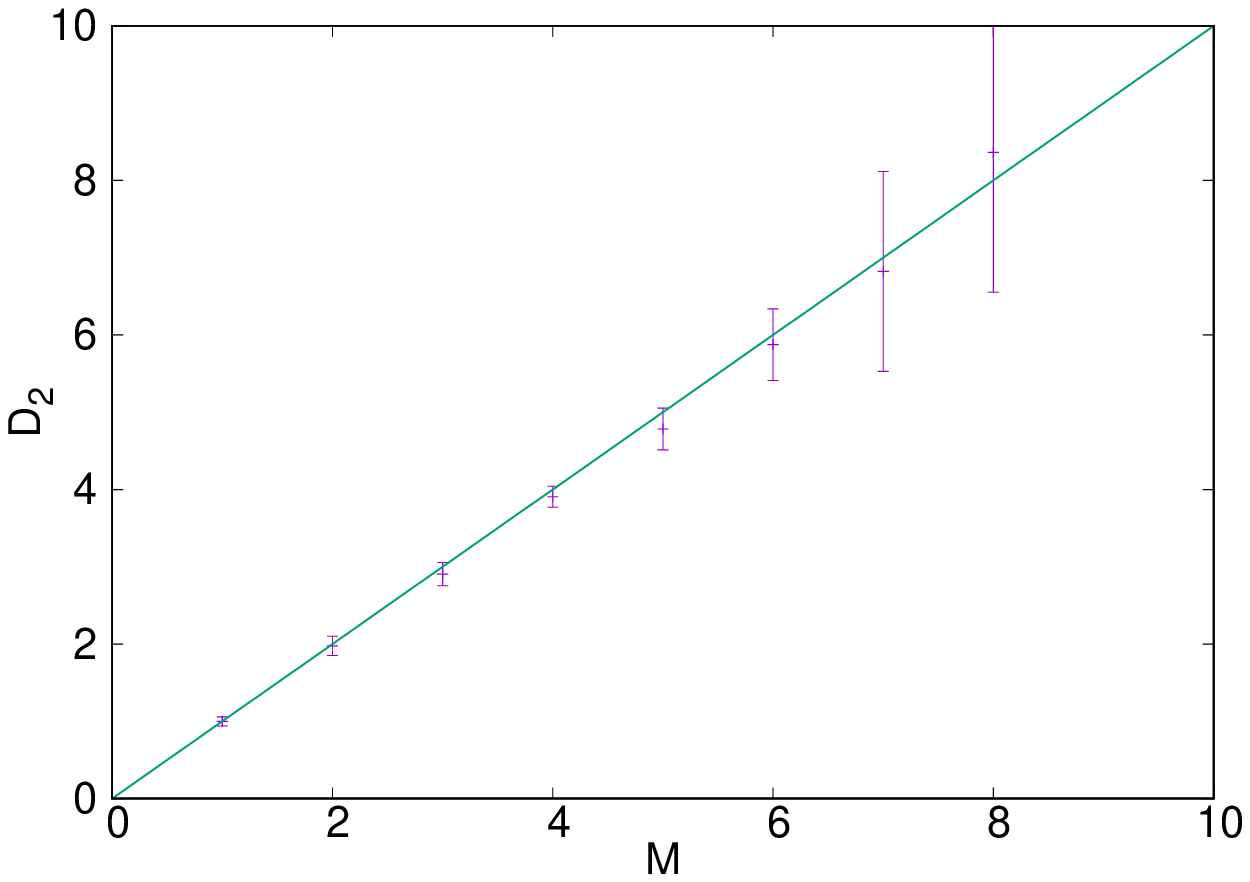}
 \end{tabular}
\caption{Top panels: Unfolded spectra for \textit{XMM-Newton} ObsIDs 0677980201 (top-left) and 0700381301 (top-right) of IGR~J17091-3624. The curves with square and circle symbols represent respectively \textit{diskbb} and powerlaw contributions and the topmost solid curve in each panel represents the overall spectrum. 
Bottom panels: Variation of correlation dimension as a function of embedding dimension with errors for \textit{XMM-Newton} ObsIDs 0677980201 (bottom-left) and 0700381301 (bottom-right) of IGR~J17091-3624. The solid diagonal line in each panel represents the ideal stochastic curve. 
}
\label{fig:fig2}
\end{figure*}
\begin{figure*}
\centering
 \begin{tabular}{@{}cc@{}}
  \includegraphics[scale=0.20, angle=-90]{chandra05.eps} &
  \includegraphics[scale=0.20, angle=-90]{chandra06.eps}
  \\
  \includegraphics[scale=0.45, angle=0]{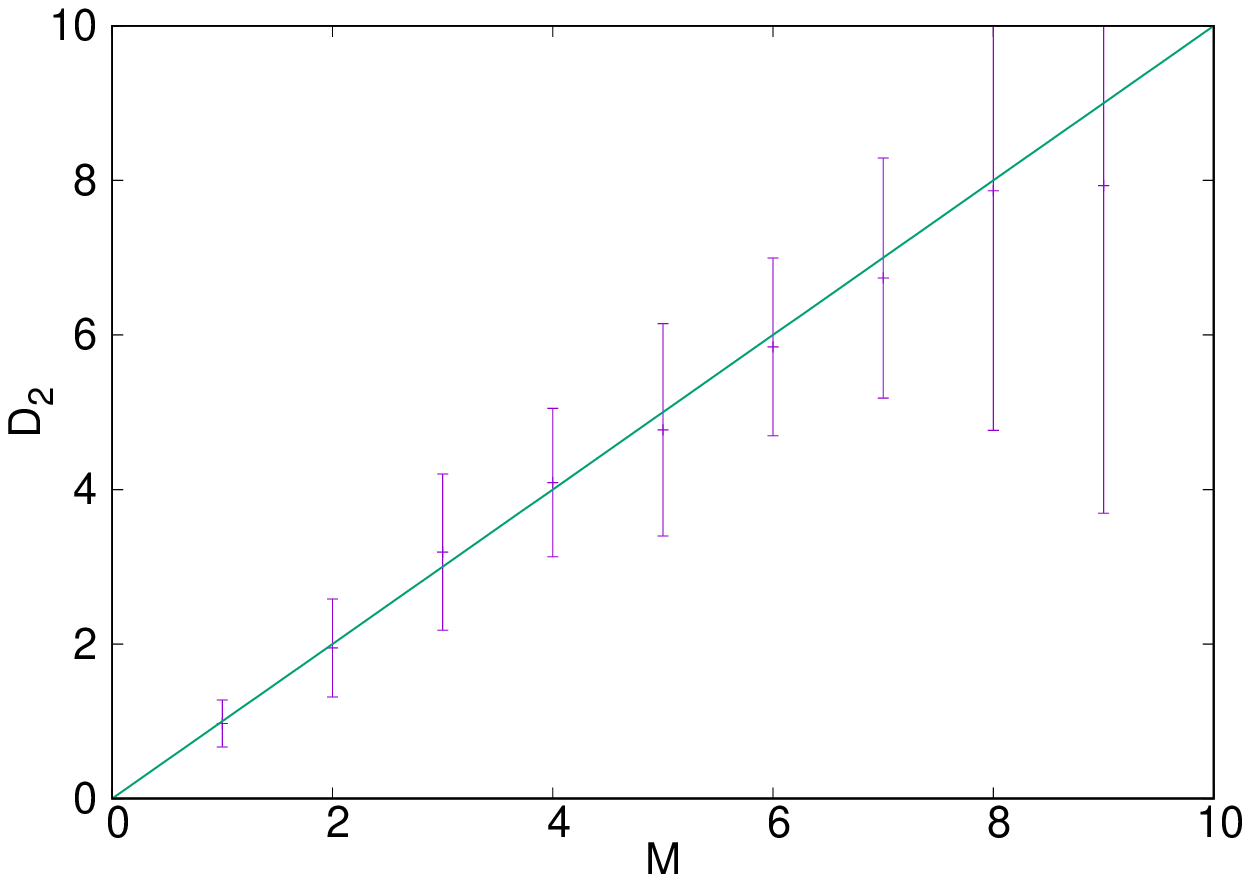} &
  \includegraphics[scale=0.45, angle=0]{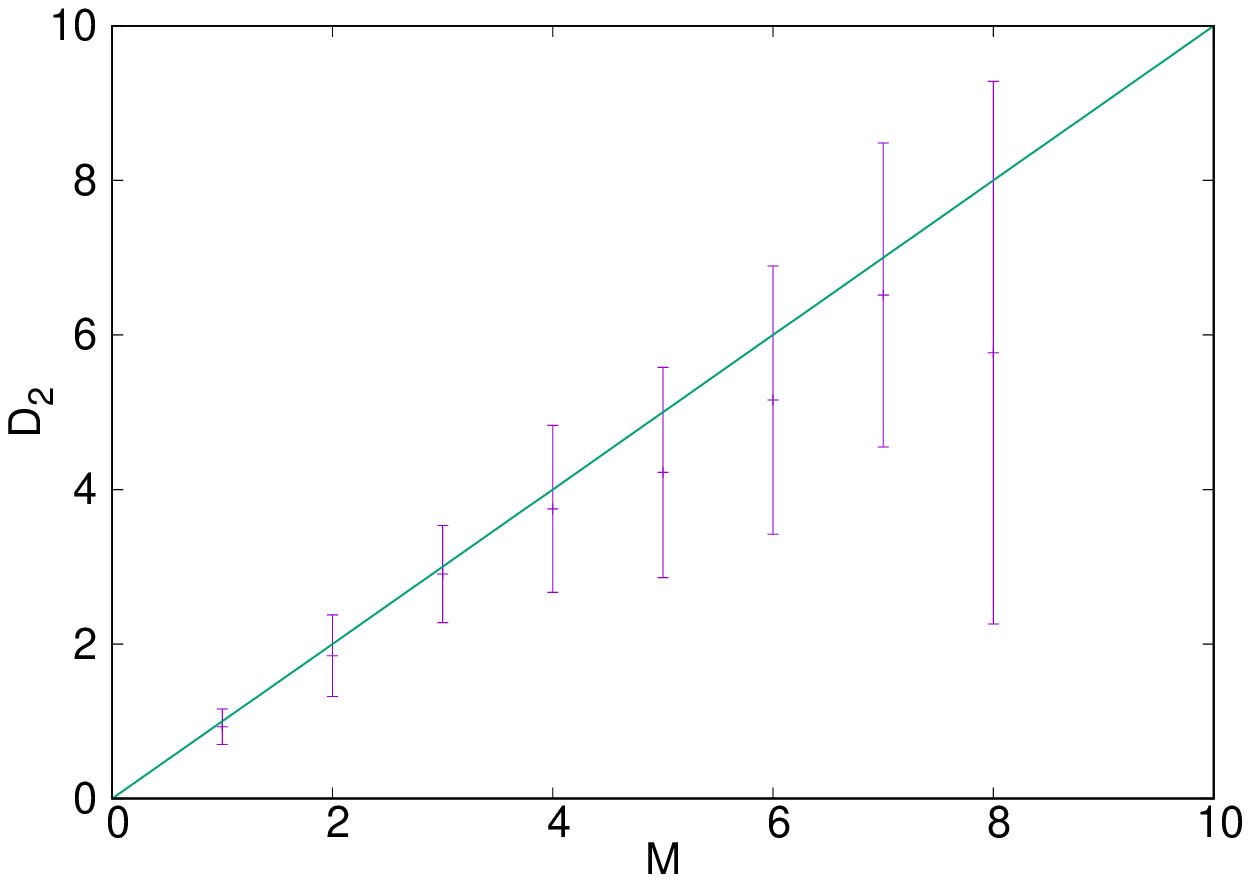}
 \end{tabular}
\caption{Top panels: Unfolded spectra for \textit{Chandra} ObsIDs 12405 (top-left) and 12406 (top-right) of IGR~J17091-3624. The curves with square and circle symbols represent respectively \textit{diskbb} and powerlaw contributions and the topmost solid curve in each panel represents the overall spectrum. 
Bottom panels: Variation of correlation dimension as a function of embedding dimension with errors for \textit{Chandra} ObsIDs 12405 (bottom-left) and 12406 (bottom-right) of IGR~J17091-3624. The solid diagonal line in each panel represents the ideal stochastic curve. 
}
\label{fig:fig3}
\end{figure*}

\begin{figure*}
\begin{tabular}{@{}cc@{}}
\includegraphics[scale=0.38,angle=0]{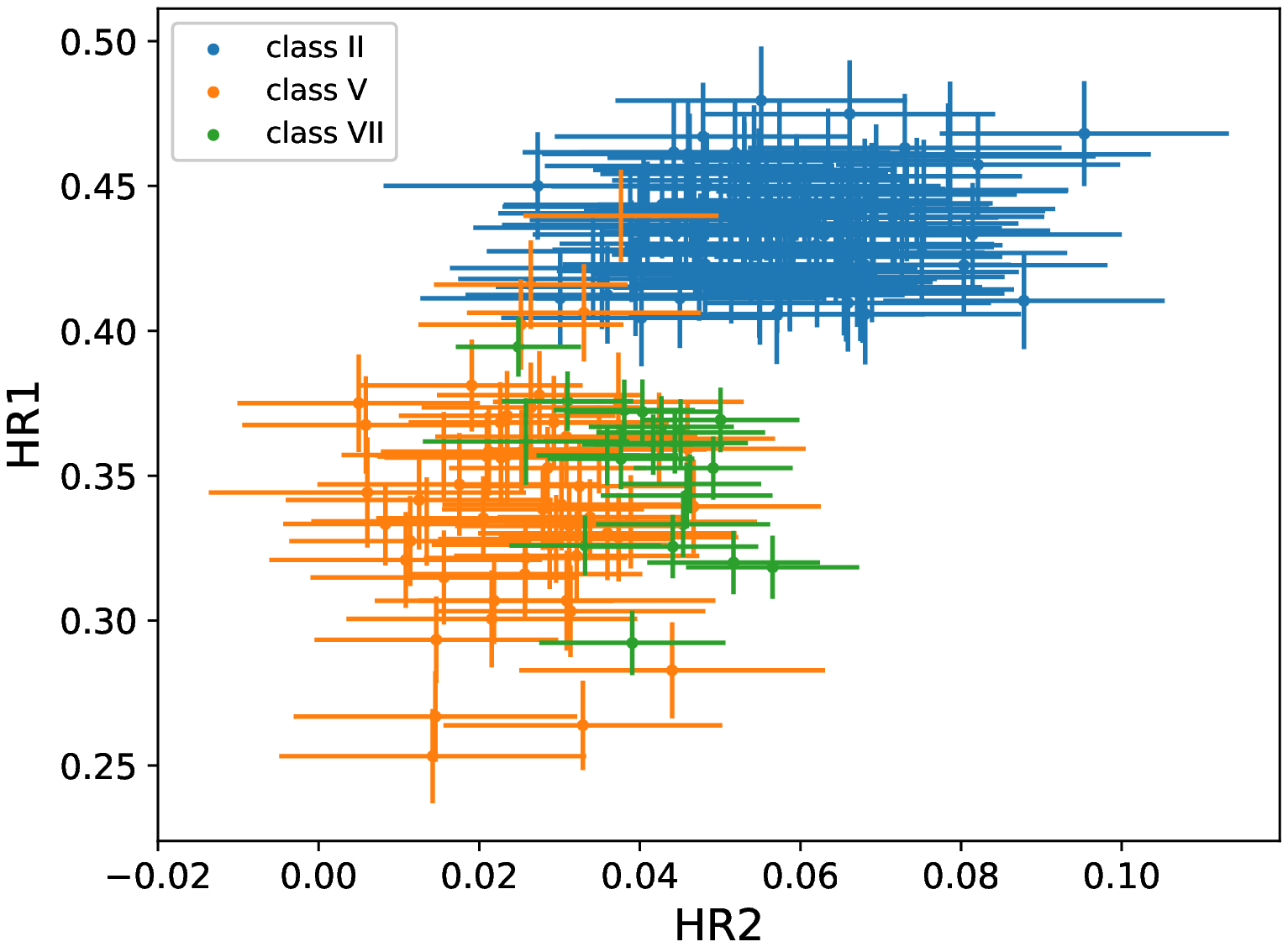}
\includegraphics[scale=0.38,angle=0]{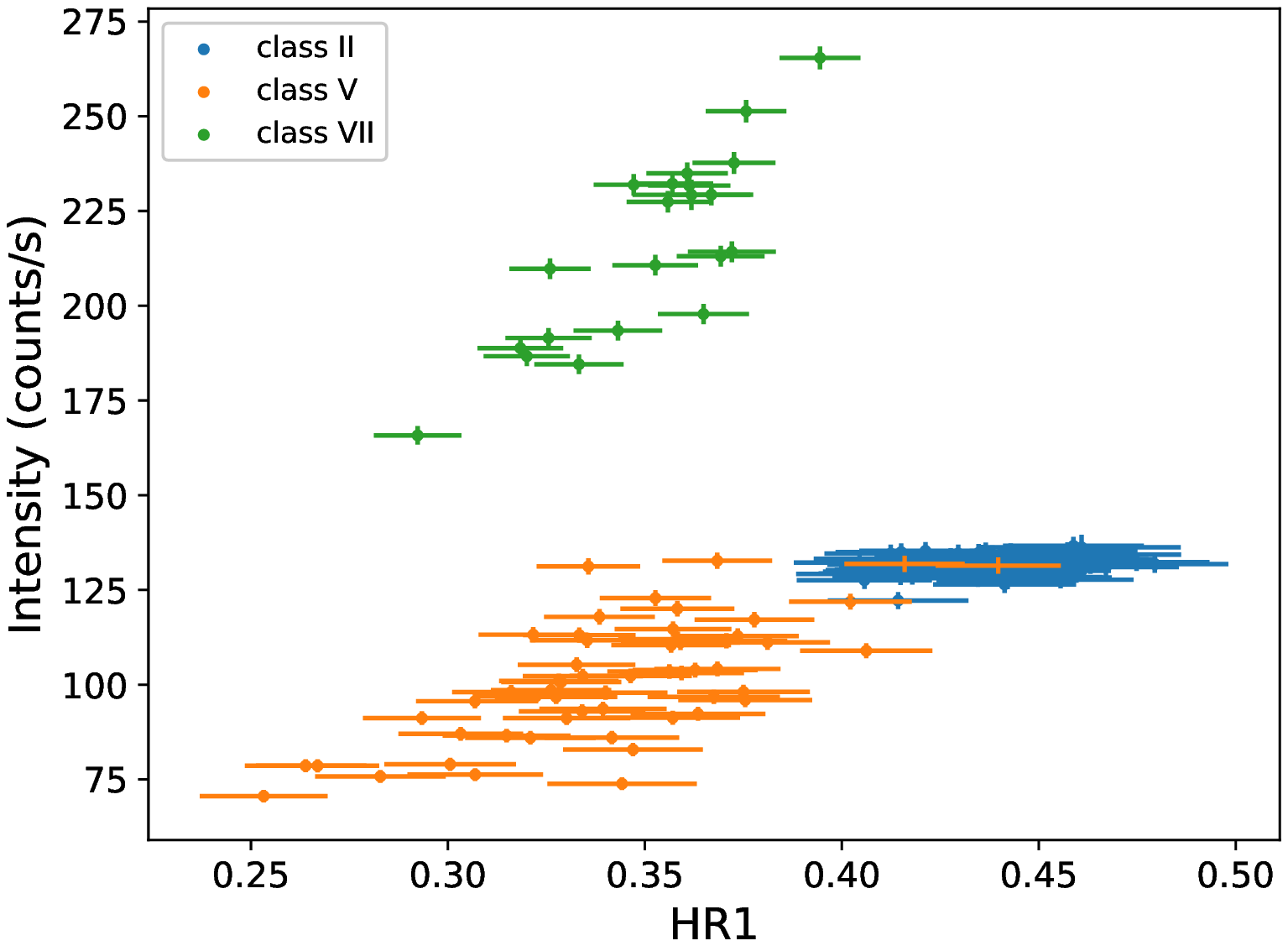} 
\includegraphics[scale=0.38,angle=0]{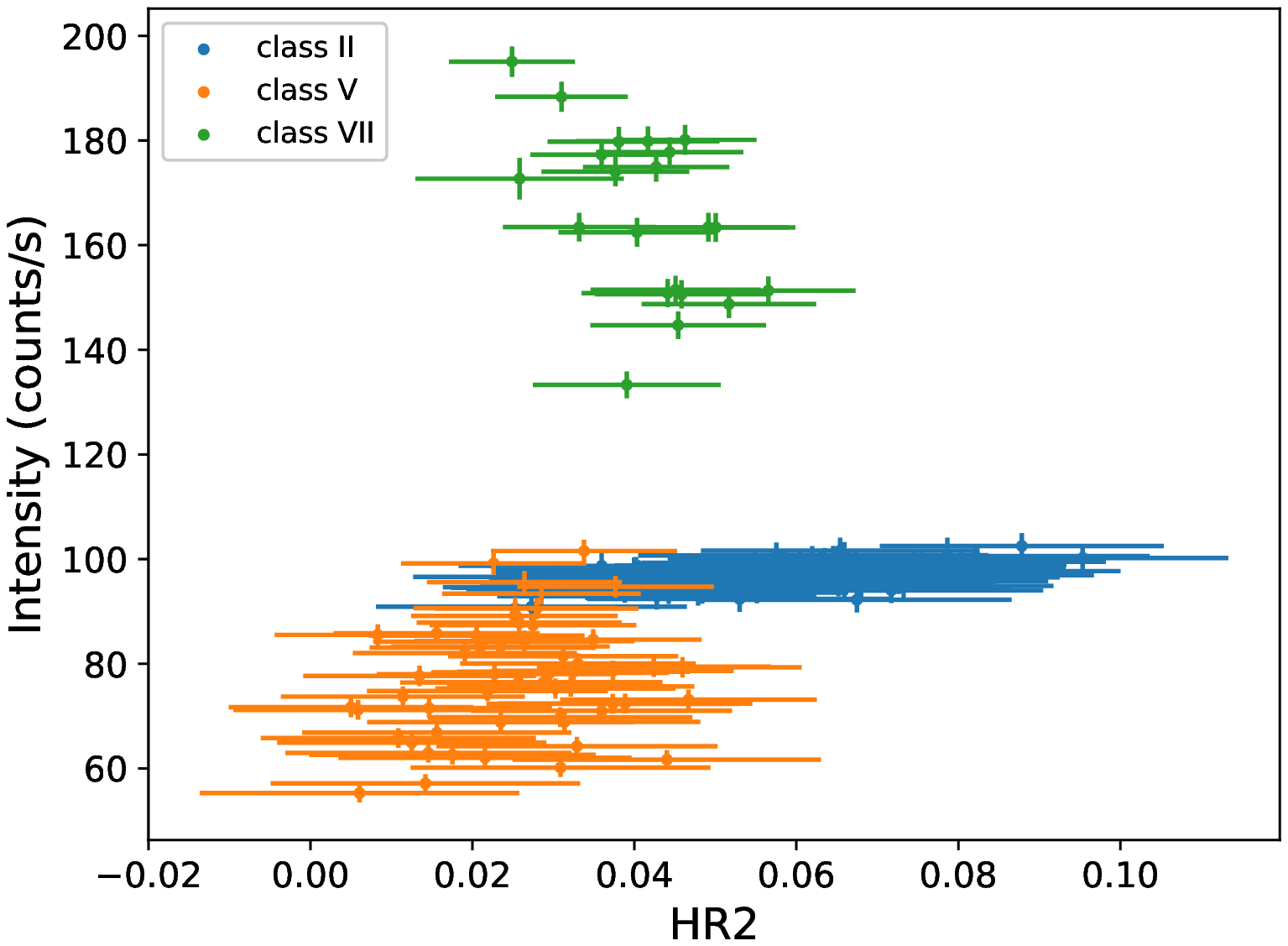}
\end{tabular}
\caption{Left panel: Colour-Colour diagrams for classes II, V and VII of \textrm{IGR~J17091-3624}
from \textit{RXTE}, given in Table~1. Middle and Right panels: Hardness-Intensity diagrams for the same classes. See text for details.
}
\label{fig:fig4}
\end{figure*}

\begin{table*}
\caption{\textit{XMM-Newton} observations}
\label{tab:tab2}
\small
\resizebox{\linewidth}{!}{
\begin{tabular}{cccccccccccccccccccccc}\\
\hline
\hline
 ObsID & IGR~J17091 class & behaviour & \textit{diskbb} & PL & GA & SI & $\chi^{2}/dof$ & state & \textit{diskbb} $T_{in}$ & $F\,\times10^{-10}$ \\
\hline
\hline
0677980201 & IV & S & $72.1$ & $27.9$ & $-$ & $1.41^{+0.18}_{-0.18}$ & $1.04$ & DD & $1.16\pm{0.01}$ & $17.562$\\
0700381301 & After \textit{RXTE} & S & $39.4$ & $60.6$ & $-$ & $1.31^{+0.02}_{-0.02}$ & $1.09$ & PD & $0.73\pm{+0.04}$ & $0.938$\\
\hline
\hline

\end{tabular}
}
{Columns:- 
1: Observational identification number.
2: \textrm{IGR~J17091-3624} temporal class.
3: The behaviour of the system (F: low correlation/fractal dimension.
S: Poisson noise like stochastic).
4: $\%$ of multi-colour disc blackbody component.
5: $\%$ of powerlaw component.
6: $\%$ of Gaussian line component (XSPEC model gauss).
7. Powerlaw photon spectral index.
8: Reduced $\chi^2$.
9: Spectral state (DD: disc dominated; PD: powerlaw dominated).
10: \textit{diskbb} temperature in units of $keV$.
11: Total flux in the $0.5-8$ keV in units of $erg\,cm^{-2}\,s^{-1}$.
}
\end{table*}
\begin{table*}
\caption{\textit{Chandra} observations}
\label{tab:tab3}
\small
\resizebox{\linewidth}{!}{
\begin{tabular}{cccccccccccccccccccccc}\\
\hline
\hline
 ObsID &  IGR~J17091 class & behaviour & \textit{diskbb} & PL & GA & SI & $\chi^{2}/dof$ & state & \textit{diskbb} $T_{in}$ & $F\,\times10^{-10}$ \\
\hline
\hline
12405 & VII & S & $63.8$ & $31.7$ & $-$ & $0.51^{+0.95}_{-0.95}$ & $1.318$ & DD & $1.19^{+0.08}_{-0.08}$ & $12.760$\\
12406 & IX & S & $50.5$ & $49.5$ & $-$ & $0.58^{+0.80}_{-0.80}$ & $1.016$ & PD & $1.53^{+0.15}_{-0.15}$ & $1.220$\\
\hline
\hline

\end{tabular}
}
{Columns:- 
1: Observational identification number.
2: \textrm{IGR~J17091-3624} temporal class.
3: The behaviour of the system (F: low correlation/fractal dimension;
S: Poisson noise like stochastic).
4: $\%$ of multi-colour disc blackbody component.
5: $\%$ of powerlaw component.
6: $\%$ of Gaussian line component (XSPEC model gauss).
7: Powerlaw photon spectral index.
8: Reduced $\chi^2$.
9: Spectral state (DD: disc dominated; PD: powerlaw dominated).
10: \textit{diskbb} temperature in units of $keV$.
11: Total flux in the $0.5-8$ keV in units of $erg\,cm^{-2}\,s^{-1}$.
}
\end{table*}

\section{Observation and data reduction}
The data we use in our analysis were obtained during the $2011-2013$ outburst of \textrm{IGR~J17091-3624} from \textit{RXTE}, \textit{XMM-Newton} and \textit{Chandra}. 
To allow for easy and efficient cross verification, we use the same data reported by \citet{2017MNRAS.468.4748C}. In all cases, we use single orbit data for each observation ID (ObsID). This is to avoid the need to interpolate over data gaps that may result from merging observations across orbits, as this may adversely affect results from the CI method.
\subsection{\textit{RXTE}}
We analyse data from the proportional counter array \citep[PCA;][]{1996SPIE.2808...59J}. We create background subtracted spectra and lightcurves following standard procedure on standard 2 and the science event data (Good Xenon). In Table \ref{tab:tab1}, we show the list of ObsIDs that we use. 
We choose the energy range $3-25\,\mathrm{keV}$ for our \textit{RXTE} spectral analysis.
For our timing analysis, we bin the science event lightcurves to $0.125\,\mathrm{s}$, 
unless stated otherwise, and the standard 2 lightcurves to $32\,\mathrm{s}$. In order to probe the hardness variability using a model independent approach over the duration of the outburst, we further consider three energy bands: A($2-6\,\mathrm{keV}$), B($6-16\,\mathrm{keV}$) and C($16-60\,\mathrm{keV}$), using the $32\,\mathrm{s}$ binned standard 2 lightcurves.

\subsection{\textit{XMM-Newton}}
We use two \textit{XMM-Newton} observations of \textrm{IGR~J17091-3624} with ObsIDs $0677980201$ and $0700381301$ (given in Table~\ref{tab:tab2}) in our analysis. For both observations, we use EPIC-pn data obtained in \textit{timing~mode} 
and we generate lightcurves with a time bin of $5\,\mathrm{s}$. We employ the \textit{XMM-Newton} Science Analysis System (SAS v.16.1.0) package for data reduction with updated Current Calibration Files (CCFs). We generate spectra and lightcurves from the product of the meta-task XMMEXTRACTOR.
\subsection{\textit{Chandra}}
We use two \textit{Chandra} observations with ObsIDs 12405 and 12406 
(given in Table~\ref{tab:tab3}) out of seven 
observations of \textrm{IGR~J17091-3624} made during the period $2011-2013$. We choose the observations to coincide with the period when the source was being monitored by \textit{RXTE}. We employ \textrm{CIAO} version 4.8 (Fruscione et al. 2006) in our analysis of the data. While ObsID 12405 was observed in the Continuous Clocking mode, 12406 was observed in Time Exposure mode. The lightcurves
are generated with time bin of $2\,\mathrm{s}$.
\begin{table*}
\caption{IGR~J17091-3624: Poisson noise to $rms$ variation using time bin of $0.125\,\mathrm{s}$}
\label{tab:tab4}
\resizebox{\linewidth}{!}{
\begin{tabular}{cccccccccccccccccccccc}\\
\hline
\hline
ObsID & class & GRS~1915-like class &  $\langle S \rangle$ & $rms$ & $\langle PN \rangle$ & $\langle PN \rangle/rms$ &  Behaviour \\
\hline
\hline
96420-01-01-00 & I & $\chi$  & $13.78$ & $4.69$ & $3.71$ & $0.79$ & S \\
\hline
96420-01-11-00 & II & $\phi$  & $15.95$ & $5.29$ & $3.99$ & $0.75$ & S \\
\hline
96420-01-04-01 & III & $\nu$  & $22.75$ & $6.51$ & $4.77$ & $0.73$ & S \\
\hline
96420-01-05-00 & IV & $\rho$  & $29.50$ & $9.47$ & $5.43$ & $0.57$ & S \\
\hline
96420-01-06-03 & V & $\mu$  & $12.38$ & $7.55$ & $3.52$ & $0.47$ & S \\
\hline
96420-01-09-00 & VI & $\lambda$  & $30.84$ & $9.79$ & $5.55$ & $0.57$ & S \\
\hline
96420-01-18-05 & VII & $none$  & $26.78$ & $16.33$ & $5.17$ & $0.32$ & S \\
\hline
96420-01-19-03 & VIII & $none$  & $30.83$ & $16.55$ & $5.55$ & $0.33$ & S \\
\hline
96420-01-35-02 & IX & $\gamma$  & $28.28$ & $9.61$ & $5.31$ & $0.55$ & S \\
\hline
\hline
\end{tabular}
}
\\
{Columns:- 
1: {\textit RXTE} ObsID from which the data has been taken. 
2: Temporal class based on the classification by Court et al. (2017). 
3: Corresponding \textrm{GRS~1915+105}-like classification. 
4: Average counts in the lightcurves $\langle S \rangle$. 
5: The $rms$ variation in the lightcurve counts. 
6: The expected Poisson noise variation, $\langle {PN} \rangle=\langle {S} \rangle ^{1/2}$. 
7: The ratio of the expected Poisson noise to the actual $rms$ variation. 
8: The behaviour of the system.}
\end{table*}
\section{Results from analyses}
In our probe of the properties of IGR~J17091-3624 during its 2011-2013 outburst, we follow closely the classification described in \citet{2017MNRAS.468.4748C} based on data from \textit{RXTE}, \textit{XMM-Newton} and \textit{Chandra}.\\

\subsection{\textit{RXTE}}
The details of the \textit{RXTE} observations that we use in our analysis are provided in Tables \ref{tab:tab1}, \ref{tab:tab4} and \ref{tab:tab5}. In Fig. \ref{fig:fig1}, we show the spectral fit for three of the typical
observations, exhibiting powerlaw dominated (PD), similar contributions to \textit{diskbb} and powerlaw 
(D-P) and \textit{diskbb} dominated (DD) spectra in the upper panels and their respective correlation dimensions in the lower panels. 
The detailed procedure of the computation of correlation dimension $D_2$ and nomenclature for spectral
behaviour were reported in the previous literature (e.g. \citealt{2010ApJ...708..862K,2017MNRAS.466.3951A}),
part of which are only briefly reported in \S 2.
Interestingly, the time series of all the classes primarily show stochastic behaviour with unsaturated 
correlation dimension, as given by 
Table~1. This is somewhat counter intuitive, when  \textrm{IGR~J17091-3624} 
is generally believed to be similar to \textrm{GRS~1915+105} with several common temporal
classes to each other, whereas \textrm{GRS~1915+105} exhibits some classes to be fractal (or 
may be chaotic) and some other to be stochastic. Note also that class
I exhibits high state emitting hard photons.
As flux decreases in class II, it still remains in hard state leading 
the flow to in low-hard state.

We note however that by increasing time bin above $0.125\,\mathrm{s}$, 
classes V and VIII seem to deviate from pure stochasticity and hence may 
exhibit higher dimensional fractal/chaotic signatures, which we term as NS, 
following previous work (\citealt{2004ApJ...609..313M}), as also indicated
in Tables~1 and 5 (see Section 5). On the other hand, increasing time bin decreases the number of lightcurve data points concurrently, hence the result is to be taken with caution (the approach is best if the number of points in a lightcurve is of the order of a few tens of thousand). 
\subsection{\textit{XMM-Newton}}
In Fig. \ref{fig:fig2} we show the fitted spectra (upper panels) from the two \textit{XMM-Newton} observations and the respective correlation dimensions are shown in the lower panels. We consider the energy range $0.5-8.0\,\mathrm{keV}$ for spectral analysis. ObsID 0677980201 has been categorised Class IV variability properties by \citet{2017MNRAS.468.4748C}. The second observation (ObsID: 070038130) was carried out after decommissioning \textit{RXTE} and, hence, cannot be compared with \textit{RXTE} data. Like \textit{RXTE} data, all 
of them show stochastic behaviour. However, the spectra for ObsID 0677980201 appears to be DD, 
which is on the other hand equivalent to Class IV of \textit{RXTE} (ObsID 96420-01-05-00) 
exhibiting PD. This
discrepancy is due to the different ranges of energy of respective satellites. If we analyse
the ObsID 0677980201 of \textit{XMM-Newton} for the energy range $3-8$ keV, then it indeed 
shows also PD spectra like \textit{RXTE}, with however a very high PL contribution.
This may be due to the satellite's higher spectral resolution than \textit{RXTE}, 
also indicating most of the thermal photons to be arising from lower energy regime. In fact,
in $3-8$ keV the corresponding ObsID 96420-01-05-00 of \textit{RXTE} show D-P, with
however indecent reduced $\chi^2$ ($\sim 3$). This again
argues against spectral resolving power of \textit{RXTE} particularly in the
lower energy regime. 

\subsection{\textit{Chandra}}
The spectral fits and correlation dimensions of two \textit{Chandra} data considered in 
our analysis are shown in Fig. \ref{fig:fig3}. We consider the energy range $0.5-8.0\,\mathrm{keV}$ for spectral analysis. According to \citet{2017MNRAS.468.4748C}, ObsID 12405 resembles a Class VII lightcurve. 
 ObsID 12406 on the other hand is considered to show properties similar to those seen in Class IX. Both the underlying time series exhibit 
stochasticity, conforming with \textit{RXTE} and \textit{XMM-Newton} findings as well
as equivalent \textrm{GRS~1915+105} Class $\rho$. The ObsID 12405 with Class VII also shows
DD spectra, similar to its \textit{RXTE} counter-part. However, ObsID 12406 with Class IX
exhibits D-P, where its \textit{RXTE} counter-part shows DD as shown in Table~1. This may
be due to higher spectral resolution of \textit{Chandra}. Indeed, the corresponding reduced 
$\chi^2$ for ObsID 12406 is much better than that of ObsID 96420-01-35-02.

\subsection{Model independent analysis}
In order to understand non-subjective properties, we
study the colour-colour diagram (CD) and hardness-intensity diagram (HID) of each class using \textit{RXTE} PCU standard $2$ data. The 
hard colour is defined as HR2=C/A and the soft colour as HR1=B/A. In Fig. \ref{fig:fig4} we show the CDs and corresponding HIDs for all nine \textit{RXTE} classes. The colour values provide a straight-forward and non-subjective way to track spectral evolution.

\citealt{2017MNRAS.468.4748C} probed loop directions in the HID of \textrm{IGR~J17091-3624}. The CD in the left panel of Fig. \ref{fig:fig4} confirms that indeed class II contains much harder photon flux compared to classes V and VII as seen in the model spectra in Table \ref{tab:tab1}. This is also evident in the middle and right panels of Fig. \ref{fig:fig4} which also show that class VII has the highest photon count rate.
Their analyses reveal evolution of two branches in the CD as the source transits from one class to another during the duration of the outburst.

\section{Significance of Observed Stochasticity}
As stated already, {\rm IGR~J17091-3624} is about a factor of $20$ fainter than {\rm GRS~1915+105} in the $2-25\,\mathrm{keV}$ energy range and, as such, the effect of noise in the observed data should not be underestimated. To quantify this plausible effect, we compute the Poisson noise level in our data and subsequently generate surrogate data to determine the significance on our results for {\rm IGR~J17091-3624}.
\subsection{Poisson Noise Effect}
Poisson noise is associated with photon counting devices -- which follow Poisson distribution -- due basically to the quantised nature of light particles and the independence of photon detections. It grows relatively weaker for higher photon counts.

\citet{2004ApJ...609..313M}, in their nonlinear time series analysis of {\rm GRS~1915+105}, estimated the ratios of expected Poisson noise $\langle PN \rangle$ to the $rms$ variation of counts for each of the representative ObsIDs reported by \citet{2000A&A...355..271B}. The Poisson noise is defined as $\langle PN \rangle \equiv \langle S \rangle ^{1/2}$, where $\langle S \rangle$ is the average photon counts in each observation. They found that classes which show chaotic behaviour have the smallest ratios $\langle PN \rangle /rms \lesssim0.05$, while classes showing stochasticity have higher ratios $\langle PN \rangle/rms \gtrsim0.2$. They therefore posit that low-dimensional chaotic signatures in the temporal behaviours of black holes may be detectable only when Poisson fluctuations are much smaller than the variability. They however did not rule out the possibility that there indeed may be a stochastic component to the variability that dominates for some temporal classes of {\rm GRS~1915+105}, seen for most BHXBs (e.g. Cyg~X-1). Interestingly, in an attempt to investigate the mechanism generating the bursts in {\rm GRS~1915+105}, {\rm IGR~J17091-3624} and {\rm MXB~1730-335}, \citet{2018A&A...612A..33M} noted that although the signal to noise ratio 
 for the \textit{RXTE} data of {\rm IGR~J17091-3624} is low due to low source counts, the observed variability is intrinsic to the source. Further, they found 
only marginal similarities between the burst properties observed in {\rm GRS~1915+105} and {\rm IGR~J17091-3624}. Therefore, our found dissimilarities in 
nonlinear time series properties between {\rm GRS~1915+105} and  
{\rm IGR~J17091-3624} may be in order.

Nevertheless, to quantify the plausible significance of Poisson fluctuations in the \textit{RXTE} data of {\rm IGR~J17091-3624}, we also compute the ratios of the expected Poisson noise $\langle PN \rangle$ to the actual $rms$ variations of counts for each ObsID, shown in Tables \ref{tab:tab4} and \ref{tab:tab5}. All cases show $\langle PN \rangle/rms \gtrsim0.2$. This implies plausibly that Poisson noise effects may not be unconnected with the stochasticity seen in all temporal classes of {\rm IGR~J17091-3624}. This effect however may be pronounced due to the much lower brightness of {\rm IGR~J17091-3624} compared to {\rm GRS~1915+105}.
\begin{table*}
\caption{IGR~J17091-3624: Poisson noise to $rms$ variation using time bin of $0.5\,\mathrm{s}$}
\label{tab:tab5}
\resizebox{\linewidth}{!}{
\begin{tabular}{cccccccccccccccccccccc}\\
\hline
\hline
ObsID & class & GRS~1915-like class &  $\langle S \rangle$ & $rms$ & $\langle PN \rangle$ & $\langle PN \rangle/rms$ &  Behaviour & $nmsd$ \\
\hline
\hline
96420-01-01-00 & I & $\chi$  & $55.12$ & $9.68$ & $7.42$ & $0.77$ & S & 0.15 \\
\hline
96420-01-11-00 & II & $\phi$  & $63.82$ & $10.79$ & $7.99$ & $0.74$ & S & 1.00 \\
\hline
96420-01-04-01 & III & $\nu$  & $90.96$ & $16.58$ & $9.54$ & $0.57$ & S & 0.59 \\
\hline
96420-01-05-00 & IV & $\rho$  & $117.98$ & $30.87$ & $10.86$ & $0.35$ & S & 2.18 \\
\hline
96420-01-06-03 & V & $\mu$  & $49.53$ & $26.18$ & $7.04$ & $0.27$ & S/NS & 0.61 \\
\hline
96420-01-09-00 & VI & $\lambda$  & $123.34$ & $31.92$ & $11.11$ & $0.35$ & S & 0.95 \\
\hline
96420-01-18-05 & VII & $none$  & $107.12$ & $60.60$ & $10.35$ & $0.17$ & S & 0.51 \\
\hline
96420-01-19-03 & VIII & $none$  & $123.34$ & $59.97$ & $11.10$ & $0.18$ & S/NS & 0.77 \\
\hline
96420-01-35-02 & IX & $\gamma$  & $113.11$ & $30.65$ & $10.63$ & $0.35$ & S & 0.39 \\
\hline
\hline
\end{tabular}
}
\\
{Columns:- 
1: {\textit RXTE} ObsID from which the data has been taken. 
2: Temporal class based on the classification by Court et al. (2017). 
3: Corresponding \textrm{GRS~1915+105}-like classification. 
4: Average counts in the lightcurves $\langle S \rangle$. 
5: The $rms$ variation in the lightcurve counts. 
6: The expected Poisson noise variation, $\langle {PN} \rangle=\langle {S} \rangle^ {1/2}$. 
7: The ratio of the expected Poisson noise to the actual $rms$ variation. 
8: The behaviour of the system.
9: Normalised mean sigma deviation, $nmsd$.}
\end{table*}
\begin{figure}
\includegraphics[scale=0.65]{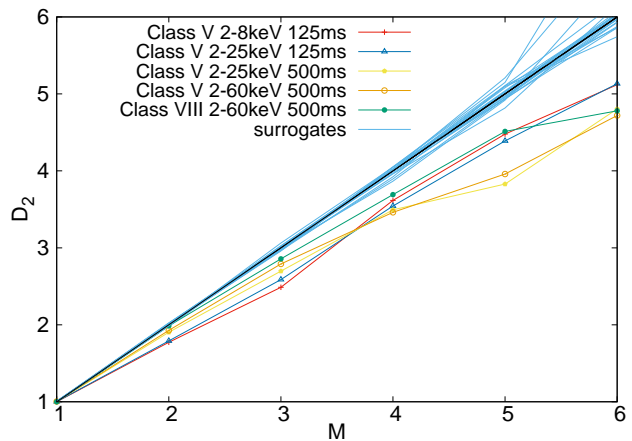}
\caption{Variation of correlation dimension $D_{2}$ as a function of embedding dimension $M$ for classes V (ObsID 96420-01-06-03) and VIII (ObsID 96420-01-19-03). The different sets of energy bins in keV and the time bins in 
ms are indicated by various symbols. The lines without points are from surrogate data 
(see \S 5.2) and the straight diagonal black line represents the behaviour of an ideal stochastic system.}
\label{fig:simd2}
\end{figure}
To mitigate against this possible effect, we further optimise both the energy band and the time bins. We carry out the analyses with increasing time bins 
from $0.125\,\mathrm{s}$ up to $1.0\,\mathrm{s}$ and energy bins of $2-8\,\mathrm{keV}$, $2-25\,\mathrm{keV}$, $2-60\,\mathrm{keV}$ and $8-25\,\mathrm{keV}$. 
The ratio $\langle PN \rangle/rms$ decreases by a few factors with increasing time bin, 
however the effect of reducing energy bin is only marginal. Figure 
\ref{fig:simd2} indicates that with increasing time bin, $D_{2}$ deviates 
from pure stochastic nature in classes V and VIII especially. This deviation is highlighted as nonstochasticity (NS) in Table \ref{tab:tab5}.
$\langle PN \rangle/rms$ values for time bin of $0.5\,\mathrm{s}$ in $2-60\,\mathrm{keV}$ are also shown in Table \ref{tab:tab5}, which decrease with respect to those for $0.125\,\mathrm{s}$ reported in Table \ref{tab:tab4}.

\subsection{Surrogate data analysis}
To place a greater level of confidence on any detected non-trivial structures in a time series, 
surrogate data analysis is normally applied. This has proven to be a very powerful method for differentiating 
signatures indicative of intrinsic fractal nature from coloured noise in time series, as shown 
in Fig. \ref{fig:simd2}. The basic idea is to 
formulate a null hypothesis that the data has been generated by a stationary linear stochastic process and 
then by comparing results for the data with appropriate realisations of surrogate data, one attempts to 
reject this hypothesis. Surrogate data sets are normally generated such that they have the same distribution 
and power spectrum as the original data. One of the standard methods used for generating surrogates is the 
Iterative Amplitude Adjusted Fourier Transform (IAAFT) method \citep{PhysRevLett.77.635, SCHREIBER2000346}, 
which is an improvement over the Amplitude Adjusted Fourier Transform (AAFT) technique \citep{THEILER199277}. 
We apply the scheme in this paper using the TISEAN package \citep{doi:10.1063/1.166424}. More detailed 
description of this method can be found in \citet{2006ApJ...643.1114M} among others.

We use the normalised mean sigma deviation ($nmsd$), proposed by \citet{HARIKRISHNAN2006137}, 
to quantify the differences in the discriminating measure between the data and the surrogates. This quantity 
is defined by
\begin{eqnarray}
nmsd^{2}=\frac{1}{M_{max}-1}\mathop{\sum^{M_{max}}}_{M=M_{min}}{\left(\dfrac{D_{2}(M)-\langle D_{2}^{surr}(M) \rangle}{\sigma_{SD}^{surr}(M)}\right)},
\end{eqnarray}
where $M_{min}$ and $M_{max}$ are the minimum and maximum embedding dimensions respectively for which the 
analysis is carried out, $D_{2}^{surr}(M)$ is the $D_{2}$ value for surrogates at a given $M$, 
$\langle D_{2}^{surr}(M) \rangle$ is the average of $D_{2}^{surr}(M)$ and $\sigma_{SD}^{surr}(M)$ is the 
standard deviation of $D_{2}^{surr}(M)$. \citet{HARIKRISHNAN2006137} showed that a value of $nmsd<3.0$ 
indicates either white noise or coloured noise domination in the data and the null hypothesis cannot 
be rejected in such a case. On the other hand, when $nmsd>3$, the data may be considered 
distinguishable from its surrogates.

For better statistics, in our computation of $nmsd$, we use the lightcurves of bin size 
$0.5\,\mathrm{s}$ and generate $19$ surrogates for each of the ObsIDs representing the nine temporal 
classes of {\rm IGR~J17091-3624}. This is because, as shown by \citet{SCHREIBER2000346}, $19$ surrogate time 
series are required for one-sided tests for a minimal significance requirement of $95\%$. In all cases, 
we fix $M_{min}=4$ in the estimation of $nmsd$ to exclude possible artefacts in the data at lower values 
of $M$. This is more justified since deterministic/fractal properties are designated based on $D_2$ 
at increasing $M$ and are preserved at higher values of 
$M$. The last column of Table \ref{tab:tab5} shows $nmsd$ values obtained for all $9$ temporal classes 
of {\rm IGR~J17091-3624}.

As discussed in \S5.1, a few of the classes tend to show signatures of plausible higher dimensional 
chaos with higher time bins. Typical cases are classes V and VIII, shown in Fig. \ref{fig:simd2} as they can 
be seen to deviate from pure stochasticity at higher values of $M$. The behaviour of these two classes are 
referred to S/NS in Table \ref{tab:tab5}. However, corresponding $nmsd$ values indicate that these deviations may 
not be significant. Longer, continuous observations with more modern X-ray telescopes providing 
more data points will be crucial to probe this better.
   
\section{Discussions and conclusion}
An accretion flow around a compact object is a nonlinear general relativistic system involving magnetohydrodynamics. Earlier, at least some of the accretion flows were argued to be similar to a Lorenz system \citep{2004ApJ...609..313M, 2004AIPC..714...48M, 2006ApJ...643.1114M, 2010ApJ...708..862K}, but contaminated by Poisson noise. 

The large variability seen in \textrm{GRS~1915+105} was interpreted by \citet{1997ApJ...479L.145B} to be the result of rapid removal and replenishment of matter in the hot accretion flow, which is driven by thermal-viscous instability in the disc. Since \textrm{IGR~J17091-3624} has been shown to mimic the variability properties seen in \textrm{GRS~1915+105}, it is natural to expect that the unique variability seen in both sources may be driven by the same physics \citep[see, e.g.,][]{2012MNRAS.422.3130C}.

\citet{2018MNRAS.476.1581A}, in their study of the connection between observed spectra and nonlinear properties of \textrm{GRS~1915+105}, interpreted the accretion modes of the source in terms of four accretion classes, namely, Keplerian disc, slim disc, advection dominated accretion flow and general advective accretion flow. They also argued that accretion rate must play an important role in transition from one accretion mode to others. We therefore attempt to interpret our analysis results for \textrm{IGR~J17091-3624} in the line with this picture.

\subsection{General features obtained in time series}
For each of the $9$ temporal classifications of \textrm{IGR~J17091-3624}, we probe their underlying nonlinear time series properties through correlation integral method -- a method that has been established over the years as an efficient technique for understanding low-dimensional chaotic properties in time series. This can be typically quantified by the variation of correlation dimension $D_{2}$ against the embedding dimension $M$. In this way, the value of $D_{2}$ determines the effective number of differential equations describing the dynamics of the system and the variation of $D_{2}$ with $M$ reveals the nonlinear dynamical properties of the system. The system is considered deterministic if initially $D_{2}$ increases linearly with $M$ until it reaches a certain value and saturates. On the other hand, the system is stochastic if for all $M$, $D_{2}\approx M$  \citep[see e.g.,][]{2004ApJ...609..313M,2004AIPC..714...48M,2017MNRAS.466.3951A}.

Unlike in \textrm{GRS~1915+105}, where some of the temporal classes show signatures of low-dimensional fractal (or F) while others show signatures consistent with stochasticity (or S), for \textrm{IGR~J17091-3624} we find all $9$ classes to be stochastic in their time series behaviour as shown in Table \ref{tab:tab1}. Equally all the lightcurves obtained with \textit{XMM-Newton} and \textit{chandra} show signatures consistent with stochasticity.
\subsection{General features obtained in spectral analysis}
We fit the energy spectra for all observations using XSPEC version $12.9.1$. Errors are quoted at the one sigma confidence level for all parameters. The two main components are the multi-colour disc blackbody model or \textit{diskbb} \citep[see, e.g.,][]{1984PASJ...36..741M,1986ApJ...308..635M} and a powerlaw model or \textit{PL}. The multiplicative model \textit{tbabs} is used to account for X-ray absorption by the inter-stellar medium (ISM), with the equivalent hydrogen column density $n_{H}$ fixed at $6.28\times10^{21}\,\mathrm{{cm}^{-2}}$ \citep{2005A&A...440..775K}. For \textit{Chandra} and \textit{XMM-Newton} data, $n_{H}$ was allowed to vary from $6.28\times10^{21}\,\mathrm{{cm}^{-2}}$ to about a factor of 2 higher for acceptable fit parameters. The
quantity ${\chi}^{2}/dof$ for classes I, IV and VIII gets improved considerably, from $70/45$, $59/45$ and $59/45$ to $48/43$, $45/43$ and $56/43$ respectively, with the inclusion of Gaussian line features fixed at $6.4\,\mathrm{keV}$ (representing Fe-$\textrm{K}_{\alpha}$ lines). The fit to the \textit{RXTE} data of class IX gets worsened with the inclusion of Gaussian line at $6.4\,\mathrm{keV}$ with ${\chi}^{2}/dof$ changing from $71/45$ to $75/43$. All the other classes give acceptable spectral fits and do not require the inclusion of Gaussian line features. As evident in Table \ref{tab:tab1}, four of the nine classes (VI, VII, VIII, IX) exhibit \textit{diskbb} dominated spectrum (DD), four (I, II, III, IV) show \textit{PL} dominated spectrum (PD), while one reveals almost equal contributions from \textit{diskbb} and \textit{PL}. We define a state to be PD if the ratio of contributions from \textit{PL} to \textit{diskbb} is $\gtrsim4/3$, and DD if the ratio of contributions from \textit{diskbb} to \textit{PL} is $\gtrsim4/3$. In terms of spectral properties, classes I, III, IV and IX are similar to those of their GRS~1915-like counterparts discussed in \citet{2018MNRAS.476.1581A}. While class II represents a typical low-hard state with the lowest flux, class I may depict the hard intermediate state, being powerlaw dominated with a high flux \citep{2016ASSL..440...61B}.

Of the four extreme combinations of accretion flows described by \cite{2018MNRAS.476.1581A} (namely: F \& HS; S \& HS; F \& LH and S \& LH), primarily two (S \& HS and S \& LH) can be defined for \textrm{IGR~J17091-3624}. 
In the line with this classification, high-soft (HS) state showing S in a time series implies quasi-spherical but radiation trapped, optically thick flow revealing incoherent flux in time, otherwise known as slim disc and is usually associated with highest accretion rates \citep[e.g.][]{1988ApJ...332..646A}. On the other hand, low-hard (LH) state showing S in a time series is consistent with a flow that is optically thin, puffed-up and 
quasi-spherical, where random cooling processes result in incoherent flux making it stochastic and possessing 
strong radial advection and is typically associated with significantly low accretion rates, 
called general advective accretion flow (GAAF: \citealt{2010MNRAS.402..961R}). 
We posit that in \textrm{IGR~J17091-3624}, depending on accretion rate, the flow switches from one accretion 
regime to another. At the high accretion rate (or the regime of highest accretion rate among four possible
above mentioned cases, corresponding to ``S \& HS"), the flow mimics a slim disc which is radiation dominated and geometrically thicker than a Keplerian disc, which is seen in the temporal classes VI, VII, VIII and IX. At significantly low accretion rate 
(but not at the regime of lowest accretion rate, which corresponds to ``F \& LH", see
\citealt{2018MNRAS.476.1581A}),
the flow becomes hot, radiatively inefficient, optically thin and with significant advection, namely GAAF, 
as seen in the temporal classes I, II, III and IV. Indeed, Table~\ref{tab:tab1} shows that on average the 
former four classes exhibit higher luminosity than latter ones, with the exception in class I, which 
has been explained above. This favours our apparently heuristic argument. However, the flow must pass through in between
the Keplerian regime with intermediate accretion rates. This seems to be revealed in 
classes V and VIII at higher time bins with reduced Poisson noise, based on the analysis discussed 
in \S5.1. Emergence of non-stochasticity in time series with DD spectral (HS)
state indicates the flow to be between slim disc and Keplerian disc regimes (pure fractal nature would 
argue for a pure Keplerian flow). On the other hand, emergence of non-stochasticity along with D-P state 
favours the idea of the flow to be between Keplerian and GAAF regimes. 
Note that a DD state corresponds to higher accretion rate than a D-P state, which is indeed
revealed from the respective flux reported in Table~\ref{tab:tab1}. Nevertheless, as discussed in \S5.2, all the classes including V and VIII exhibit $nmsd<3$, arguing 
against any deterministic feature.
Hence, longer time series data (which is not
available for data sets under consideration at $1\,\mathrm{s}$ bin) are required to confirm this hypothesis. The future X-ray missions that can observe continuously for much longer duration 
may provide such required lightcurves and hence time series.
Note that classes identified 
as slim discs should in principle be fitted with \textit{diskpbb} rather than \textit{diskbb} model in XSPEC.
Hence, we have subsequently cross-verified that the underlying spectral 
fits indeed are consistent with \textit{diskpbb} model and they are indeed consistent with
slim discs. 

Therefore, while the presence of plausibly significant level of Poisson noise in our data 
poses problem for comparison of the nonlinear timing properties of 
\textrm{IGR~J17091-3624} vis-a-vis \textrm{GRS~1915+105}, they seem to be following similar track, at 
least partially. 
This is because underlying low-dimensional fractal signature may have been suppressed, as the presence of noise can significantly affect $D_{2}-M$ relation and suppress any signatures of low dimensional chaos in a system \citep[see, e.g.,][]{2004ApJ...609..313M,2006ApJ...643.1114M}. However, the stochastic nature may indeed be inherent in 
the lightcurves of \textrm{IGR~J17091-3624}. If this is so, it implies that although \textrm{IGR~J17091-3624} and \textrm{GRS~1915+105} show remarkable similarity in their exotic variability patterns, the underlying physics responsible for this may not be very same.

\section*{Acknowledgements}

This work was partly supported by a project supported by Department of Science and Technology 
with Grant No. DSTO/PPH/BMP/1946 (EMR/2017/001226). The authors thank 
the referee for his valuable 
comments and suggestions which prompted us to rethink some preliminary conclusion and
further analysis that improved the quality of the work.
The thanks are also due to H. Sreehari for useful help in relation to \textit{RXTE} data reduction.

\bsp	
\label{lastpage}

\begin{thebibliography}{}
\makeatletter
\relax
\def\mn@urlcharsother{\let\do\@makeother \do\$\do\&\do\#\do\^\do\_\do\%\do\~}
\def\mn@doi{\begingroup\mn@urlcharsother \@ifnextchar [ {\mn@doi@}
  {\mn@doi@[]}}
\def\mn@doi@[#1]#2{\def\@tempa{#1}\ifx\@tempa\@empty \href
  {http://dx.doi.org/#2} {doi:#2}\else \href {http://dx.doi.org/#2} {#1}\fi
  \endgroup}
\def\mn@eprint#1#2{\mn@eprint@#1:#2::\@nil}
\def\mn@eprint@arXiv#1{\href {http://arxiv.org/abs/#1} {{\tt arXiv:#1}}}
\def\mn@eprint@dblp#1{\href {http://dblp.uni-trier.de/rec/bibtex/#1.xml}
  {dblp:#1}}
\def\mn@eprint@#1:#2:#3:#4\@nil{\def\@tempa {#1}\def\@tempb {#2}\def\@tempc
  {#3}\ifx \@tempc \@empty \let \@tempc \@tempb \let \@tempb \@tempa \fi \ifx
  \@tempb \@empty \def\@tempb {arXiv}\fi \@ifundefined
  {mn@eprint@\@tempb}{\@tempb:\@tempc}{\expandafter \expandafter \csname
  mn@eprint@\@tempb\endcsname \expandafter{\@tempc}}}

\bibitem[\protect\citeauthoryear{{Abramowicz}, {Czerny}, {Lasota}  \&
  {Szuszkiewicz}}{{Abramowicz} et~al.}{1988}]{1988ApJ...332..646A}
{Abramowicz} M.~A.,  {Czerny} B.,  {Lasota} J.~P.,   {Szuszkiewicz} E.,  1988,
  \mn@doi [\apj] {10.1086/166683}, \href
  {http://adsabs.harvard.edu/abs/1988ApJ...332..646A} {332, 646}

\bibitem[\protect\citeauthoryear{{Adegoke}, {Rakshit}  \&
  {Mukhopadhyay}}{{Adegoke} et~al.}{2017}]{2017MNRAS.466.3951A}
{Adegoke} O.,  {Rakshit} S.,   {Mukhopadhyay} B.,  2017, \mn@doi [\mnras]
  {10.1093/mnras/stw3320}, \href
  {http://adsabs.harvard.edu/abs/2017MNRAS.466.3951A} {466, 3951}

\bibitem[\protect\citeauthoryear{{Adegoke}, {Dhang}, {Mukhopadhyay}, {Ramadevi}
   \& {Bhattacharya}}{{Adegoke} et~al.}{2018}]{2018MNRAS.476.1581A}
{Adegoke} O.,  {Dhang} P.,  {Mukhopadhyay} B.,  {Ramadevi} M.~C.,
  {Bhattacharya} D.,  2018, \mn@doi [\mnras] {10.1093/mnras/sty263}, \href
  {http://adsabs.harvard.edu/abs/2018MNRAS.476.1581A} {476, 1581}

\bibitem[\protect\citeauthoryear{{Altamirano} \& {Belloni}}{{Altamirano} \&
  {Belloni}}{2012}]{2012ApJ...747L...4A}
{Altamirano} D.,  {Belloni} T.,  2012, \mn@doi [\apjl]
  {10.1088/2041-8205/747/1/L4}, \href
  {http://adsabs.harvard.edu/abs/2012ApJ...747L...4A} {747, L4}

\bibitem[\protect\citeauthoryear{{Altamirano} et~al.,}{{Altamirano}
  et~al.}{2011}]{2011ApJ...742L..17A}
{Altamirano} D.,  et~al., 2011, \mn@doi [\apjl] {10.1088/2041-8205/742/2/L17},
  \href {http://adsabs.harvard.edu/abs/2011ApJ...742L..17A} {742, L17}

\bibitem[\protect\citeauthoryear{{Ardito}, {Ricciardi}, {Massaro}, {Mineo}  \&
  {Massa}}{{Ardito} et~al.}{2017}]{2017IJNLM..88..142A}
{Ardito} A.,  {Ricciardi} P.,  {Massaro} E.,  {Mineo} T.,   {Massa} F.,  2017,
  \mn@doi [International Journal of Non Linear Mechanics]
  {10.1016/j.ijnonlinmec.2016.10.017}, \href
  {https://ui.adsabs.harvard.edu/abs/2017IJNLM..88..142A} {88, 142}

\bibitem[\protect\citeauthoryear{{Belloni} \& {Motta}}{{Belloni} \&
  {Motta}}{2016}]{2016ASSL..440...61B}
{Belloni} T.~M.,  {Motta} S.~E.,  2016, in {Bambi} C.,  ed.,  Astrophysics and
  Space Science Library Vol. 440, Astrophysics of Black Holes: From Fundamental
  Aspects to Latest Developments. p.~61 (\mn@eprint {arXiv} {1603.07872}),
  \mn@doi{10.1007/978-3-662-52859-4_2}

\bibitem[\protect\citeauthoryear{{Belloni}, {M{\'e}ndez}, {King}, {van der
  Klis}  \& {van Paradijs}}{{Belloni} et~al.}{1997}]{1997ApJ...479L.145B}
{Belloni} T.,  {M{\'e}ndez} M.,  {King} A.~R.,  {van der Klis} M.,   {van
  Paradijs} J.,  1997, \mn@doi [\apjl] {10.1086/310595}, \href
  {http://adsabs.harvard.edu/abs/1997ApJ...479L.145B} {479, L145}

\bibitem[\protect\citeauthoryear{{Belloni}, {Klein-Wolt}, {M{\'e}ndez}, {van
  der Klis}  \& {van Paradijs}}{{Belloni} et~al.}{2000}]{2000A&A...355..271B}
{Belloni} T.,  {Klein-Wolt} M.,  {M{\'e}ndez} M.,  {van der Klis} M.,   {van
  Paradijs} J.,  2000, \aap, \href
  {http://adsabs.harvard.edu/abs/2000A%26A...355..271B} {355, 271}

\bibitem[\protect\citeauthoryear{{Capitanio}, {Del Santo}, {Bozzo}, {Ferrigno},
  {De Cesare}  \& {Paizis}}{{Capitanio} et~al.}{2012}]{2012MNRAS.422.3130C}
{Capitanio} F.,  {Del Santo} M.,  {Bozzo} E.,  {Ferrigno} C.,  {De Cesare} G.,
   {Paizis} A.,  2012, \mn@doi [\mnras] {10.1111/j.1365-2966.2012.20834.x},
  \href {http://adsabs.harvard.edu/abs/2012MNRAS.422.3130C} {422, 3130}

\bibitem[\protect\citeauthoryear{{Castro-Tirado}, {Brandt}  \&
  {Lund}}{{Castro-Tirado} et~al.}{1992}]{1992IAUC.5590....2C}
{Castro-Tirado} A.~J.,  {Brandt} S.,   {Lund} N.,  1992, \iaucirc, \href
  {http://adsabs.harvard.edu/abs/1992IAUC.5590....2C} {5590}

\bibitem[\protect\citeauthoryear{{Court}, {Altamirano}, {Pereyra}, {Boon},
  {Yamaoka}, {Belloni}, {Wijnands}  \& {Pahari}}{{Court}
  et~al.}{2017}]{2017MNRAS.468.4748C}
{Court} J.~M.~C.,  {Altamirano} D.,  {Pereyra} M.,  {Boon} C.~M.,  {Yamaoka}
  K.,  {Belloni} T.,  {Wijnands} R.,   {Pahari} M.,  2017, \mn@doi [\mnras]
  {10.1093/mnras/stx773}, \href
  {http://adsabs.harvard.edu/abs/2017MNRAS.468.4748C} {468, 4748}

\bibitem[\protect\citeauthoryear{{Grassberger} \& {Procaccia}}{{Grassberger} \&
  {Procaccia}}{1983}]{1983PhRvA..28.2591G}
{Grassberger} P.,  {Procaccia} I.,  1983, \mn@doi [\pra]
  {10.1103/PhysRevA.28.2591}, \href
  {http://cdsads.u-strasbg.fr/abs/1983PhRvA..28.2591G} {28, 2591}

\bibitem[\protect\citeauthoryear{{Grinberg} et~al.,}{{Grinberg}
  et~al.}{2016}]{2016ATel.8761....1G}
{Grinberg} V.,  et~al., 2016, The Astronomer's Telegram, \href
  {http://adsabs.harvard.edu/abs/2016ATel.8761....1G} {8761}

\bibitem[\protect\citeauthoryear{Harikrishnan, Misra, Ambika  \&
  Kembhavi}{Harikrishnan et~al.}{2006}]{HARIKRISHNAN2006137}
Harikrishnan K.,  Misra R.,  Ambika G.,   Kembhavi A.,  2006, \mn@doi [Physica
  D: Nonlinear Phenomena] {https://doi.org/10.1016/j.physd.2006.01.027}, 215,
  137

\bibitem[\protect\citeauthoryear{Hegger, Kantz  \& Schreiber}{Hegger
  et~al.}{1999}]{doi:10.1063/1.166424}
Hegger R.,  Kantz H.,   Schreiber T.,  1999, \mn@doi [Chaos: An
  Interdisciplinary Journal of Nonlinear Science] {10.1063/1.166424}, 9, 413

\bibitem[\protect\citeauthoryear{{Jahoda}, {Swank}, {Giles}, {Stark},
  {Strohmayer}, {Zhang}  \& {Morgan}}{{Jahoda}
  et~al.}{1996}]{1996SPIE.2808...59J}
{Jahoda} K.,  {Swank} J.~H.,  {Giles} A.~B.,  {Stark} M.~J.,  {Strohmayer} T.,
  {Zhang} W.,   {Morgan} E.~H.,  1996, in {Siegmund} O.~H.,  {Gummin} M.~A.,
  eds,  \procspie Vol. 2808, EUV, X-Ray, and Gamma-Ray Instrumentation for
  Astronomy VII. pp 59--70, \mn@doi{10.1117/12.256034}

\bibitem[\protect\citeauthoryear{{Kalberla}, {Burton}, {Hartmann}, {Arnal},
  {Bajaja}, {Morras}  \& {P{\"o}ppel}}{{Kalberla}
  et~al.}{2005}]{2005A&A...440..775K}
{Kalberla} P.~M.~W.,  {Burton} W.~B.,  {Hartmann} D.,  {Arnal} E.~M.,  {Bajaja}
  E.,  {Morras} R.,   {P{\"o}ppel} W.~G.~L.,  2005, \mn@doi [\aap]
  {10.1051/0004-6361:20041864}, \href
  {http://cdsads.u-strasbg.fr/abs/2005A%26A...440..775K} {440, 775}

\bibitem[\protect\citeauthoryear{{Karak}, {Dutta}  \& {Mukhopadhyay}}{{Karak}
  et~al.}{2010}]{2010ApJ...708..862K}
{Karak} B.~B.,  {Dutta} J.,   {Mukhopadhyay} B.,  2010, \mn@doi [\apj]
  {10.1088/0004-637X/708/1/862}, \href
  {http://adsabs.harvard.edu/abs/2010ApJ...708..862K} {708, 862}

\bibitem[\protect\citeauthoryear{{Krimm} \& {Kennea}}{{Krimm} \&
  {Kennea}}{2011}]{2011ATel.3148....1K}
{Krimm} H.~A.,  {Kennea} J.~A.,  2011, The Astronomer's Telegram, \href
  {http://adsabs.harvard.edu/abs/2011ATel.3148....1K} {3148}

\bibitem[\protect\citeauthoryear{{Kuulkers}, {Lutovinov}, {Parmar},
  {Capitanio}, {Mowlavi}  \& {Hermsen}}{{Kuulkers}
  et~al.}{2003}]{2003ATel..149....1K}
{Kuulkers} E.,  {Lutovinov} A.,  {Parmar} A.,  {Capitanio} F.,  {Mowlavi} N.,
  {Hermsen} W.,  2003, The Astronomer's Telegram, \href
  {http://adsabs.harvard.edu/abs/2003ATel..149....1K} {149}

\bibitem[\protect\citeauthoryear{{Makishima}, {Maejima}, {Mitsuda}, {Bradt},
  {Remillard}, {Tuohy}, {Hoshi}  \& {Nakagawa}}{{Makishima}
  et~al.}{1986}]{1986ApJ...308..635M}
{Makishima} K.,  {Maejima} Y.,  {Mitsuda} K.,  {Bradt} H.~V.,  {Remillard}
  R.~A.,  {Tuohy} I.~R.,  {Hoshi} R.,   {Nakagawa} M.,  1986, \mn@doi [\apj]
  {10.1086/164534}, \href {http://adsabs.harvard.edu/abs/1986ApJ...308..635M}
  {308, 635}

\bibitem[\protect\citeauthoryear{{Maselli}, {Capitanio}, {Feroci}, {Massa},
  {Massaro}  \& {Mineo}}{{Maselli} et~al.}{2018}]{2018A&A...612A..33M}
{Maselli} A.,  {Capitanio} F.,  {Feroci} M.,  {Massa} F.,  {Massaro} E.,
  {Mineo} T.,  2018, \mn@doi [\aap] {10.1051/0004-6361/201732097}, \href
  {https://ui.adsabs.harvard.edu/abs/2018A&A...612A..33M} {612, A33}

\bibitem[\protect\citeauthoryear{{Massaro}, {Ardito}, {Ricciardi}, {Massa},
  {Mineo}  \& {D'A{\`\i}}}{{Massaro} et~al.}{2014}]{2014Ap&SS.352..699M}
{Massaro} E.,  {Ardito} A.,  {Ricciardi} P.,  {Massa} F.,  {Mineo} T.,
  {D'A{\`\i}} A.,  2014, \mn@doi [\apss] {10.1007/s10509-014-1924-9}, \href
  {https://ui.adsabs.harvard.edu/abs/2014Ap&SS.352..699M} {352, 699}

\bibitem[\protect\citeauthoryear{{Miller}, {Reynolds}, {Kennea}, {King}  \&
  {Tomsick}}{{Miller} et~al.}{2016}]{2016ATel.8742....1M}
{Miller} J.~M.,  {Reynolds} M.,  {Kennea} J.,  {King} A.~L.,   {Tomsick} J.,
  2016, The Astronomer's Telegram, \href
  {http://adsabs.harvard.edu/abs/2016ATel.8742....1M} {8742}

\bibitem[\protect\citeauthoryear{{Misra}, {Harikrishnan}, {Mukhopadhyay},
  {Ambika}  \& {Kembhavi}}{{Misra} et~al.}{2004}]{2004ApJ...609..313M}
{Misra} R.,  {Harikrishnan} K.~P.,  {Mukhopadhyay} B.,  {Ambika} G.,
  {Kembhavi} A.~K.,  2004, \mn@doi [\apj] {10.1086/421005}, \href
  {http://adsabs.harvard.edu/abs/2004ApJ...609..313M} {609, 313}

\bibitem[\protect\citeauthoryear{{Misra}, {Harikrishnan}, {Ambika}  \&
  {Kembhavi}}{{Misra} et~al.}{2006}]{2006ApJ...643.1114M}
{Misra} R.,  {Harikrishnan} K.~P.,  {Ambika} G.,   {Kembhavi} A.~K.,  2006,
  \mn@doi [\apj] {10.1086/503094}, \href
  {http://adsabs.harvard.edu/abs/2006ApJ...643.1114M} {643, 1114}

\bibitem[\protect\citeauthoryear{{Mitsuda} et~al.,}{{Mitsuda}
  et~al.}{1984}]{1984PASJ...36..741M}
{Mitsuda} K.,  et~al., 1984, \pasj, \href
  {http://adsabs.harvard.edu/abs/1984PASJ...36..741M} {36, 741}

\bibitem[\protect\citeauthoryear{{Mukhopadhyay}}{{Mukhopadhyay}}{2004}]{2004AIPC..714...48M}
{Mukhopadhyay} B.,  2004, in {Kaaret} P.,  {Lamb} F.~K.,   {Swank} J.~H.,  eds,
   American Institute of Physics Conference Series Vol. 714, X-ray Timing 2003:
  Rossi and Beyond. pp 48--51 (\mn@eprint {} {astro-ph/0402222}),
  \mn@doi{10.1063/1.1780998}

\bibitem[\protect\citeauthoryear{{Narayan} \& {Yi}}{{Narayan} \&
  {Yi}}{1994}]{1994ApJ...428L..13N}
{Narayan} R.,  {Yi} I.,  1994, \mn@doi [\apjl] {10.1086/187381}, \href
  {http://adsabs.harvard.edu/abs/1994ApJ...428L..13N} {428, L13}

\bibitem[\protect\citeauthoryear{{Pahari}, {Yadav}  \&
  {Bhattacharyya}}{{Pahari} et~al.}{2014}]{2014ApJ...783..141P}
{Pahari} M.,  {Yadav} J.~S.,   {Bhattacharyya} S.,  2014, \mn@doi [\apj]
  {10.1088/0004-637X/783/2/141}, \href
  {http://adsabs.harvard.edu/abs/2014ApJ...783..141P} {783, 141}

\bibitem[\protect\citeauthoryear{{Rajesh} \& {Mukhopadhyay}}{{Rajesh} \&
  {Mukhopadhyay}}{2010}]{2010MNRAS.402..961R}
{Rajesh} S.~R.,  {Mukhopadhyay} B.,  2010, \mn@doi [\mnras]
  {10.1111/j.1365-2966.2009.15925.x}, \href
  {http://adsabs.harvard.edu/abs/2010MNRAS.402..961R} {402, 961}

\bibitem[\protect\citeauthoryear{Schreiber \& Schmitz}{Schreiber \&
  Schmitz}{1996}]{PhysRevLett.77.635}
Schreiber T.,  Schmitz A.,  1996, \mn@doi [Phys. Rev. Lett.]
  {10.1103/PhysRevLett.77.635}, 77, 635

\bibitem[\protect\citeauthoryear{Schreiber \& Schmitz}{Schreiber \&
  Schmitz}{2000}]{SCHREIBER2000346}
Schreiber T.,  Schmitz A.,  2000, \mn@doi [Physica D: Nonlinear Phenomena]
  {https://doi.org/10.1016/S0167-2789(00)00043-9}, 142, 346

\bibitem[\protect\citeauthoryear{{Shakura} \& {Sunyaev}}{{Shakura} \&
  {Sunyaev}}{1973}]{1973A&A....24..337S}
{Shakura} N.~I.,  {Sunyaev} R.~A.,  1973, \aap, \href
  {http://adsabs.harvard.edu/abs/1973A%26A....24..337S} {24, 337}

\bibitem[\protect\citeauthoryear{Theiler, Eubank, Longtin, Galdrikian  \&
  Farmer}{Theiler et~al.}{1992}]{THEILER199277}
Theiler J.,  Eubank S.,  Longtin A.,  Galdrikian B.,   Farmer J.~D.,  1992,
  \mn@doi [Physica D: Nonlinear Phenomena]
  {https://doi.org/10.1016/0167-2789(92)90102-S}, 58, 77

\makeatother
\end{thebibliography}
\end{document}